\documentclass[twocolumn,prd,floatfix,nofootinbib,longbibliography,notitlepage,superscriptaddress]{revtex4-2}

\pdfoutput=1

\usepackage{amsmath, amssymb, amsfonts, amsthm, latexsym, mathrsfs, bbm, slashed}

\usepackage[usenames,dvipsnames]{xcolor}
\usepackage[title,titletoc]{appendix}
\usepackage{graphicx}
  
\usepackage{xifthen}
\usepackage{tensor}
\usepackage{dsfont}

\usepackage[inline]{enumitem}

\usepackage{setspace}
\usepackage[marginal, multiple]{footmisc}
 
\usepackage[T1]{fontenc}
\usepackage[utf8]{inputenc}
\usepackage{lmodern}
\usepackage[normalem]{ulem}

\usepackage[unicode, colorlinks, allcolors=blue!70!black, linktocpage, pdfusetitle]{hyperref}
\definecolor{CiteColor}{rgb}{0.55,0,0}
\hypersetup{citecolor=CiteColor}
\definecolor{RefColor}{rgb}{0,0.5,0}
\hypersetup{linkcolor=RefColor}
\usepackage[all]{hypcap}

\usepackage{orcidlink}

\usepackage{verbatim}

\usepackage{cancel}

\usepackage{microtype}

\usepackage{standalone}
\usepackage{tikz}
\usetikzlibrary{arrows.meta}
\usetikzlibrary{decorations.markings}
\ifdefined\myext
\usetikzlibrary{external}
\fi

\newcommand{\nn}{\nonumber}

\newcommand{\pd}{\partial}

\newcommand{\RR}{\mathbb{R}}
\newcommand{\LdotSeff}{\vec{L} \cdot \vec{S}_{\text{eff}}}
\newcommand{\tripleprod}{\vec{L} \cdot (\vec{S}_{1}\times \vec{S}_{2})}

\DeclareMathOperator{\sn}{sn}

\DeclareMathOperator{\am}{am}

\newcommand{\eff}{\text{eff}}

\newcommand{\J}{\mathcal{J}_5}

\newcommand{\Eff}{ {S}_{\eff}\cdot {L}}

\newcommand{\pb}[1]{\left\lbrace  #1   \right\rbrace  }

\newcommand{\figschemesetup}{%
\begin{figure}
  \includegraphics[width=\linewidth]{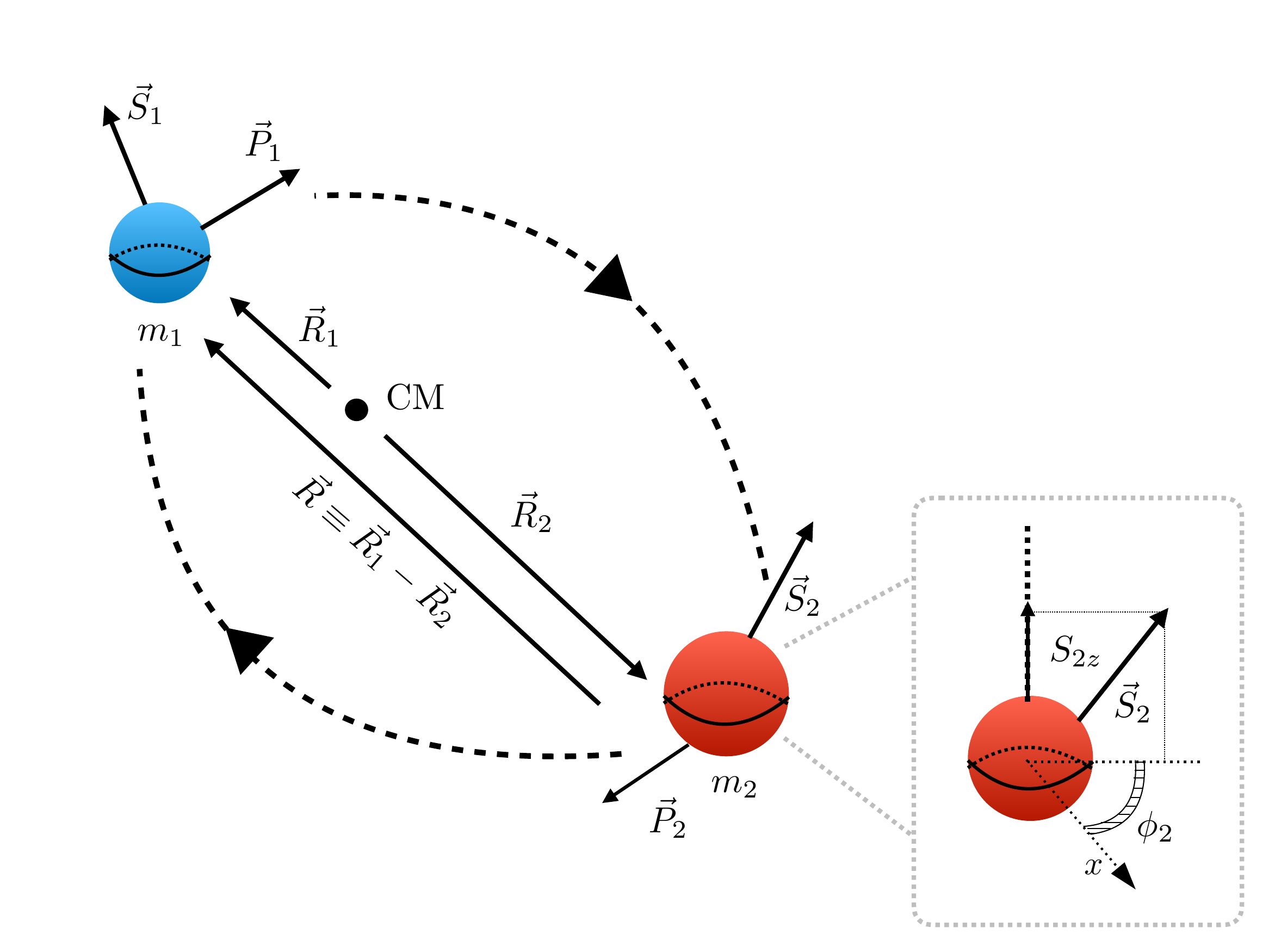}
  \caption{Schematic setup of a precessing black hole
    binary. Positions, velocities and momenta are all defined as
    Newtonian vectors built from the center-of-mass.
    \vspace{-1.em}
  }
  \label{fig:SetupFig}
\end{figure}
}

\newcommand{\figA}{%
\begin{figure}[t]
  \includegraphics[width=8cm]{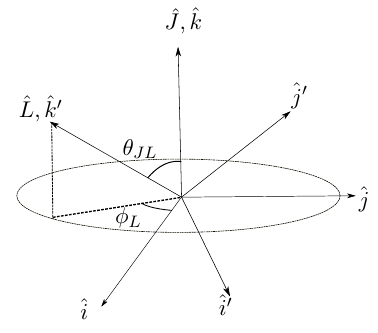}
  \caption{The non-inertial $(i'j'k')$ triad (centered around $\hat{L} \equiv \vec{L}/L$) is displayed along with the inertial $(ijk)$ triad  (centered around $\hat{J} \equiv \vec{J}/J$).} 
  \label{fig:1} 
\end{figure}
}

\newcommand{\figB}{%
\begin{figure}[tb]
  \includegraphics[width=8cm]{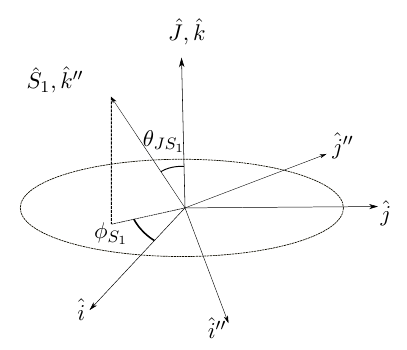}
  \caption{The second non-inertial $(i''j''k'')$ triad (centered around $\hat{S}_1 \equiv \vec{S}_1/S_1$) is displayed along with the inertial $(ijk)$ triad  (centered around $\hat{J} \equiv \vec{J}/J$).} 
  \label{fig:2} 
\end{figure}
}

\newcommand{\figC}{%
\begin{figure}
  \includegraphics[width=8cm]{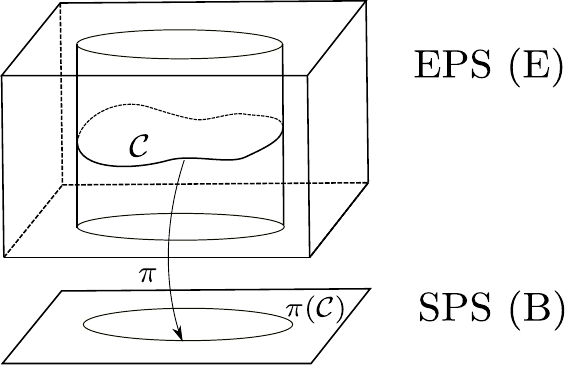}
  \caption{%
    The extended phase space $E$ can be viewed as a fiber bundle with
    projection $\pi:E\to B$ down to the standard phase space $B$.
    Both are symplectic manifolds, but the symplectic form in the EPS
    is exact. In the context of this figure, a pure vertical motion
    in the EPS corresponds to keeping the SPS coordinates fixed and changing only
    the fictitious coordinates.}
  \label{fig:fiber}
\end{figure}
}

\newcommand{\figD}{%
\begin{figure}
  \includegraphics[width=8cm]{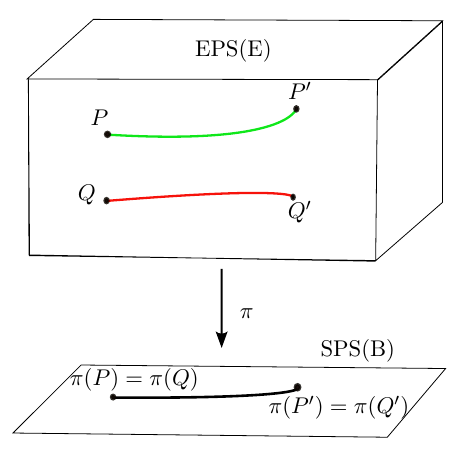}
  \caption{%
   Two EPS points ($P$ and $Q$) with the same fictitious 
   coordinates are mapped to new EPS points ($P'$ and $Q'$)
    again the the same fictitious coordinates by a flow
    under a general
    $f\left(\vec{R}, \vec{P}, \vec{S}_{1}\left(\vec{R}_{a}, 
    \vec{P}_{a}\right), \vec{S}_{2}\left(\vec{R}_{a}, \vec{P}_{a}\right)\right)$ 
    by any arbitrary amount.
    }
  \label{fig:eps_sps_1}
\end{figure}
}

\newcommand{\figE}{%
\begin{figure}
  \includegraphics[width=8cm]{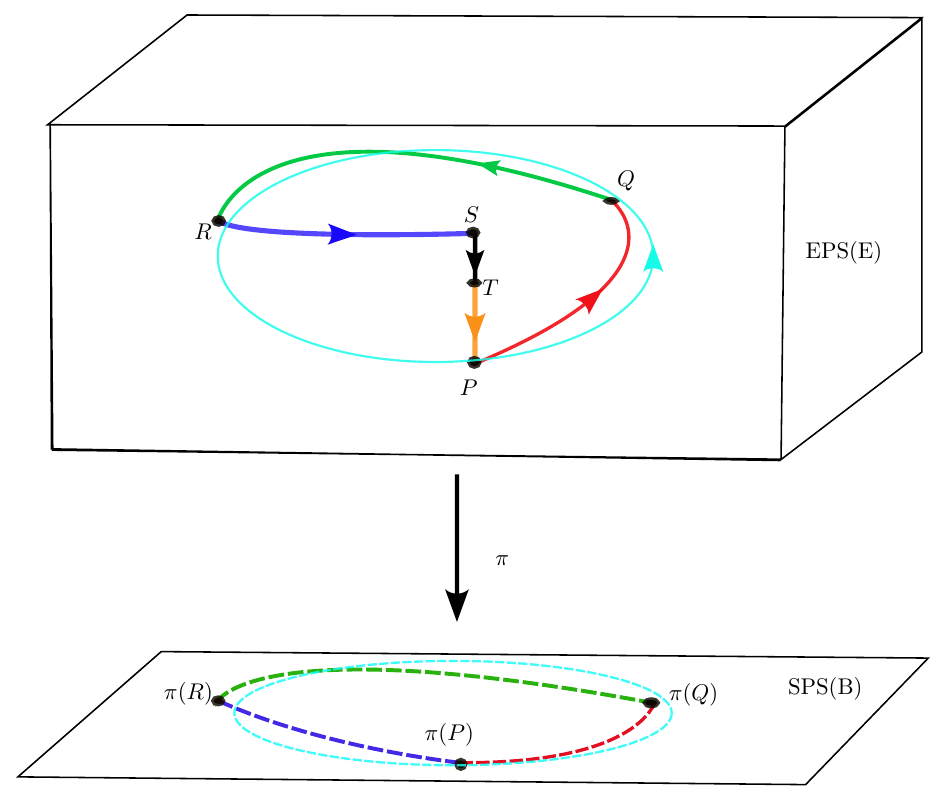}
  \caption{
  Schematic depiction of closing the loop
  in the EPS over which the fifth action integral is computed.
  This is done by flowing under $\Eff$ (red), $J^2$ (green),
  $L^2$ (blue), $S_1^2$ (black), and $S_2^2$ (orange).
  The curve in cyan is the one found by flowing
  under $\J$ in the EPS. The corresponding $\pi$ projections
  of the solid curves in the EPS is shown by broken curves in the
  SPS with the same color. The segments $ST$ and 
  $TP$ are vertical because only the fictitious variables change along these.
  }
  \label{fig:eps_sps_2}
\end{figure}
}

\begin{document}

\title{Action-angle variables of a binary black hole with arbitrary eccentricity, spins, and masses at 1.5 post-Newtonian order}

\newcommand{\UMiss}{\affiliation{Department of Physics and Astronomy, The University of Mississippi, University, MS 38677, USA}}

\newcommand{\LUX}{\affiliation{LUX, Observatoire de Paris, Université PSL, Sorbonne Université, CNRS, 92190 Meudon, France}}

\newcommand{\utec}{\affiliation{
Universidad de Ingenieria y Tecnologia – UTEC, Jr. Medrano Silva 165 - Barranco 15063, Lima, Peru
}}

\author{Sashwat Tanay\,\orcidlink{0000-0002-2964-7102}}
\email{stanay@utec.edu.pe}
\LUX
\UMiss
\utec
\author{Leo C.~Stein\,\orcidlink{0000-0001-7559-9597}}
\email{lcstein@olemiss.edu}
\UMiss
\author{Gihyuk Cho\,\orcidlink{0000-0001-8813-270X}}
\affiliation{Department of Physics and Astronomy, Seoul National University, Seoul 151-742, Korea}

\hypersetup{pdfauthor={Tanay, Cho, and Stein}}

\begin{abstract}
Accurate and efficient modeling of the dynamics of binary black
holes (BBHs) is crucial to their detection through gravitational waves
(GWs), with LIGO/Virgo/KAGRA, and LISA in the future.
Solving the dynamics of a BBH system with arbitrary parameters without simplifications
(like orbit- or precession-averaging) in closed form is one of the
most challenging problems for the GW community.
One potential approach is using canonical perturbation theory
which constructs perturbed action-angle variables from the unperturbed
ones of an integrable Hamiltonian system.
Having action-angle variables of the integrable 1.5 post-Newtonian (PN) BBH system is therefore imperative.
In this paper, we continue the work initiated by two of us in
[Tanay \textit{et al.}, \prd~103, 064066 (2021)], 
where we presented four out of five actions of a BBH system
with arbitrary eccentricity, masses, and spins, at 1.5PN order.
Here we compute the 
remaining fifth action using a novel method of
 extending the phase space by introducing
unmeasurable phase space coordinates.
We detail how to compute all the frequencies, and sketch how to
explicitly transform from the action-angle variables to the
usual positions and momenta. This analytically solves the dynamics at 1.5PN.
This lays the groundwork to analytically solve the conservative
dynamics of the BBH system with arbitrary masses, spins, and
eccentricity, at higher PN order, by using canonical perturbation theory.

\end{abstract}

\maketitle

\section{Introduction}

Laser interferometer detectors have made numerous
gravitational wave (GW) detections that have originated from compact binaries made up of
black holes (BHs) or neutron stars~\cite{ LIGOScientific:2018mvr, Abbott:2020niy, LIGOScientific:2017vwq}.
Among these detections, the predominant sources of GWs are from binary black holes (BBHs),
whose initial eccentricity is believed to be mostly radiated away by the time they enter the frequency band
of the ground-based detectors such as LIGO, Virgo, and KAGRA.
Since the upcoming LISA mission~\cite{Cutler:1997ta,Audley:2017drz} will target compact binaries
earlier in their inspiral phase compared to the ground based detectors,
incorporating eccentricity becomes more relevant. Since the
observation time for LISA sources will be much longer, it is
imperative to find accurate closed-form solutions to the binary
dynamics.

This brings us to the question of working out 
closed-form solutions of the dynamics of a \textit{generic} BBH system,
with arbitrary eccentricity, masses, and with both BHs spinning,
without special alignment.
Many such attempts have been made in the literature~\cite{tiwari_gopu,2015PhRvL.114h1103K,
2015PhRvD..92f4016G,2018PhRvD..98d4015H,2017PhRvD..95j4004C,2010PhRvD..81l4001K,2018PhRvD..98j4043K,
Konigsdorffer:2006zt, Konigsdorffer:2005sc},
but most (if not all) of them give the solution of the conservative sector of the dynamics under some
simplifying conditions such as the quasi-circular limit, equal-mass case, only one or none of the BHs spinning, with orbit-averaging, etc.
 Only recently, one of us provided a method to find the closed-form solution to
a BBH system with arbitrary eccentricity, spins, and masses at 1.5 
post-Newtonian (PN) order for the first time~\cite{cho2019analytic}
(with the 1PN part of the Hamiltonian being omitted, as it is not
complicated to handle).
The next natural question is: how can one construct the solutions at 2PN,
or is it even feasible?

This line of questioning led two of us to probe the integrability,
and therefore the existence of action-angle variables
of the BBH system at 2PN in Ref.~\cite{Tanay:2020gfb},
wherein we found that a BBH system is indeed 2PN integrable when we
applied the perturbative version of the Liouville-Arnold (LA) theorem,
due to the existence of two new 2PN constants of motion that we discovered.
Since integrability precludes chaos (which would obstruct finding closed-form solutions),
establishing integrability at 2PN
instills hope towards finding a closed-form solutions at this order.
A straightforward extension of the methods of
 Ref.~\cite{cho2019analytic} from 1.5PN to 2PN 
appears too difficult to carry out, if not impossible.
 Our hope is to use non-degenerate 
canonical perturbation theory~\cite{goldstein2013classical, jose},
which when supplied with 1.5PN action-angle variables, 
can yield 2PN action-angle variables.
If this line of work is to be pursued, the 1.5PN action-angle
variables are imperative. The calculation cannot start from a lower
PN order because the lower order (1PN) system is degenerate
in the action-angles context; this is discussed later.
We initiated the action-angle calculation in 
Ref.~\cite{Tanay:2020gfb}, where we computed
four (out of five) actions. In this paper, we compute the last action
variable, and sketch how to 
explicitly transform from the action-angle variables to the
usual positions and momenta. This basically comprises the closed-form 
solution to the  1.5PN spinning BBH dynamics.

The history of action-angle variables literature dates back
centuries. The Kepler equation presented in 1609
gives the Newtonian angle variable~\cite{goldstein2013classical},
 long before Newton proposed his
laws of motion and gravitation. Important contributions were made by
Delaunay to the action-angle formalism of the Newtonian two-body
system~\cite{goldstein2013classical} in the nineteenth century.
More recently, on the post-Newtonian front,
Damour and Deruelle gave the 1PN extension of the angle variable when they worked
out the quasi-Keplerian solution to the non-spinning
eccentric BBH system~\cite{1985AIHS...43..107D}.
Damour, Sch{\"a}fer
and Jaranowski worked out action variables at 2PN 
and 3PN ignoring the spin effects. 
Such post-Newtonian calculations make use of the work of Sommerfeld for
complex contour integration to evaluate the radial action variable~\cite{damour1988}.
Finally, Damour gave the requisite number (five) of 1.5PN constants of motion in Ref.~\cite{Damour2001},
which is required for integrability as per the LA theorem.

This paper is a natural extension to our earlier work~\cite{Tanay:2020gfb}.
We compute the remaining fifth action variable 
using a novel method of extending the phase space
by the introduction of new, unmeasurable (or fictitious)
phase space variables. We then show how to PN expand the lengthy
expression of this 1.5PN exact fifth action and 
retain the much shorter leading-order contribution.
Next we discuss how to compute all the frequencies of the system.
Then we give a clear roadmap on  how to 
compute all angle variables  of the system implicitly,
by expressing the standard phase space variables of the system
$(\vec{R}, \vec{P}, \vec{S}_1, \vec{S}_2)$ as explicit functions of
the action-angle variables.
Thereafter, we proceed towards constructing
solution to the BBH system using action-angle variables
at 1.5PN. This action-angle-based solution can be extended to 
higher PN orders via canonical perturbation theory.
Finally, in one of the appendices, we point 
out a loophole in the definition of 
PN integrability that we presented in
Ref.~\cite{Tanay:2020gfb} and also provide a simple fix.
We mention here that in a companion paper 
\cite{Samanta:2022yfe}, we implemented our action-angle based 
solution using \textsc{Mathematica}, and compared it with the
corresponding numerical solution.

The organization of this paper is as follows.
 In Sec.~\ref{setup}, we lay the conceptual foundations,
introducing the phase space (symplectic manifold) and the Hamiltonian of the system.
This includes introducing important definitions 
like those of integrability and action-angle variables.
In Sec.~\ref{sec:EPS}, we discuss the idea of extending the phase space 
by introducing new, unmeasurable phase space variables; they 
make the computation of the fifth 
action possible. In the next section, we implement these ideas to actually compute the fifth action in explicit form.
Then in Sec.~\ref{sec:lin_action}, we show 
how to PN expand this fifth action and present its shortened form.
In Sec.~\ref{sec:freq}, we finally show how to 
compute the five frequencies, the angle variables,
and construct the action-angle-based solution to the system.
Finally, we summarize our work and suggest its future extensions 
in Sec.~\ref{summary}. 
As far as appendices are concerned,
some lengthy calculations have been 
pushed to Appendix~\ref{flow_under_SeffdL},
which would have otherwise been a part of 
Sec.~\ref{sec:comp-fifth-action}.
In Appendix~\ref{SPS-action}, we prove that
our fifth action calculated in the extended phase space is also an action in the
standard phase space.
Appendix~\ref{derivative calculations} gives some 
commonly occurring derivatives that occur in the frequency calculations.
Lastly, in Appendix~\ref{fix_definition},
we  fix a loophole in
the definition of PN integrability that we
 presented in Ref.~\cite{Tanay:2020gfb}.

\section{The setup}   \label{setup}

\figschemesetup

The paper is a continuation of the research initiated
 in Ref.~\cite{Tanay:2020gfb}
and uses the same conventions, which we now briefly describe.
For an informal and pedagogical introduction to the mathematical machinery employed 
in this paper and Ref.~\cite{Tanay:2020gfb}, the reader is referred to 
the set of lecture notes at \cite{Tanay:2022rlz}.

We will study the BBH system in
the PN approximation within the Hamiltonian formalism.
The system under consideration is
schematically displayed in Fig.~\ref{fig:SetupFig}.
 We work in the center-of-mass frame with a
relative separation vector $\vec{R} \equiv \vec{r}_1 - \vec{r}_2$ between the two black holes,
and conjugate momentum $\vec{P} \equiv \vec{p}_1 = -\vec{p}_2$, where
the labels 1 and 2 indicate the two black holes, with masses $m_1$ and
$m_2$ respectively.  In Ref.~\cite{Tanay:2020gfb},
$\vec{R}_{1,2}$ and $\vec{P}_{1,2}$ were used to denote the position and momentum vectors of the
two BHs; but here we are reserving these symbols for 
to-be-introduced unmeasurable,
fictitious variables (see Sec.~\ref{sec:EPS}).
The BHs possess spin angular momenta
$\vec{S}_1$ and $\vec{S}_2$ which contribute to the total angular momentum
$\vec{J} \equiv \vec{L} + \vec{S}_1 + \vec{S}_2$, where $\vec{L}
\equiv \vec{R} \times \vec{P}$ is the
orbital angular momentum of the binary. We will frequently use the effective spin
\begin{align}
\vec{S}_{\text{eff}} &  \equiv \sigma_1 \vec{S}_1 + \sigma_2 \vec{S}_2,      \\
\sigma_1   &  \equiv 1 + \frac{3 m_2}{4 m_1}  ,    \\
\sigma_2   &  \equiv 1 + \frac{3 m_1}{4 m_2}     .    
\end{align}
The magnitude of any vector will be denoted by the same
letter used to denote the vector, but without the arrow.
Additionally, $\Eff$, and $R_a \cdot P_a$ will stand 
for $\vec{S}_{\text{eff}} \cdot \vec{L}$, 
and $\vec{R}_a \cdot \vec{P}_a$, respectively.
Einstein summation convention will be assumed unless stated otherwise.

The 1.5PN Hamiltonian that we will primarily 
be interested in is given by Eqs.~(11), (12), (13)
and (14) in Ref.~\cite{Tanay:2020gfb}, and will be denoted by $H$.
Note that $H$ in this paper is found by dropping the 2PN contribution in
the $H$ of Ref.~\cite{Tanay:2020gfb}.
The only non-vanishing Poisson brackets (PBs) between the phase space variables
$\vec{R}, \vec{P}, \vec{S}_1$, and $\vec{S}_2$ are
\begin{align}
  \left\{R^{i}, P_{j}\right\}=\delta_{j}^{i} \,,
  \quad \text{and} \quad
  \left\{S_{a}^{i}, S_{b}^{j}\right\}=\delta_{a b} \epsilon^{i j}{ }_{k} S_{a}^{k}\,,
  \label{PB1}
\end{align}
and those related by antisymmetry
\begin{align}
\{f,g \} = - \{g,f \}  .      \label{anti-symmetry}
\end{align}
Here the letters $a,b$ label the
two black holes ($a,b=1,2$), and $i,j,k$ are spatial vector indices.
The PBs are derivations, so obey the
chain rule (with $\xi^{i}$'s standing for all phase-space
variables)
\begin{equation}
\left\{f, g\left(\xi^{i}\right)\right\}=\left\{f, \xi^{i}\right\} \frac{\partial g}{\partial \xi^{i}}.          \label{chain-rule}
\end{equation}      
Eqs.~\eqref{PB1}, \eqref{anti-symmetry}, and \eqref{chain-rule}
enable us to compute the PB between any two
functions of the phase-space variables.
As usual, the evolution of any phase-space function $f$ is
given by $\dot{f} = \pb{f,H}$. With this, it can be verified that both the spin
magnitudes are constant, $\dot{S_a}  = \pb{ S_a, H }=0$.
This means that we can specify each spin vector using only two variables:
the $z$ component and the azimuthal angle of the spin vector.
This choice is particularly useful
because these two variables act like canonical
 ones. This is so because Eqs.~\eqref{PB1}, \eqref{anti-symmetry}, and
\eqref{chain-rule} imply that
\begin{align}
\left\{\phi_a , S_{b}^{z}  \right\}=  \delta_{ab}  \,.    \label{PB2}
\end{align}
This means that there are five pairs of canonically conjugate 
variables,
and a total of ten canonical phase space variables.

From a more mathematical point of view, 
Hamiltonian dynamics takes place
on a symplectic manifold $B$ which is a 
smooth manifold equipped with a closed, non-degenerate
differential 2-form $\Omega$, the symplectic form.
The orbital variables $R^{i}, P_{j}$ are 
canonical variables of the cotangent bundle
$T^{*}\RR^{3}$ (a symplectic manifold),
while each spin vector $S^{i}_{a}$ lives on the
surface of a two-sphere (also a symplectic manifold, with symplectic
form proportional to the area 2-form).
The spin vectors $\vec{S}_a$ being on the above 
spherical symplectic manifolds is consistent
with the constancy of the spin magnitudes.
The symplectic manifold $B$ which is the total phase space
of the system is the Cartesian product of the above
symplectic manifolds ($T^{*}\RR^{3}$, and the two 2-spheres).
The symplectic form on $B$ is the sum of the symplectic forms from
the three manifold factors~\cite{Tanay:2020gfb}.
In terms of canonically conjugate variables, $\Omega$ is\footnote{%
The relationship between symplectic form
and PBs is encapsulated in Eq.~(5.79) of
Ref.~\cite{jose}. The form
$\Omega$ of Eq.~\eqref{eq:sym_form}
 is consistent with the PBs of Eqs.~\eqref{PB1} and
\eqref{PB2}.}
\begin{align}
\Omega &=d P_{i} \wedge d R^{i}+d S_{1}^{z} \wedge d \phi_{1}+d S_{2}^{z} \wedge d \phi_{2}
\,  . 
\label{eq:sym_form}
\end{align}
This description of the phase space manifold using
a symplectic geometry point of view makes it clear
that each spin has only two 
degrees of freedom ($S_{a}^{z}$ and $ \phi_{a}$), rather than three
($S_a^x, S_a^y,$ and $S_a^z$).
Although $\Omega$ itself is smooth, notice that this
coordinate system is singular at the poles of each spin space.

Now we define integrable systems and action-angle variables at the
same time, re-presenting the definition given in
Ref.~\cite{fasano}.
Two quantities
$f$ and $g$ are called commuting or
``in involution'' if $\{f,g \} = 0$.
Consider a system with Hamiltonian $H$ in $2n$
canonical phase space variables
$(\vec{\mathcal{P}},\vec{\mathcal{Q}})$.  This system is integrable if
there exists a canonical transformation to coordinates
$(\vec{\mathcal{J}}, \vec{\theta})$ such that all the actions
$\mathcal{J}^{i}$ are mutually commuting, $H$ is a function
only of the actions, and all the $\vec{\mathcal{P}}$ and
$\vec{\mathcal{Q}}$ variables are $2\pi$-periodic functions of the
angle variables $\vec{\theta}$.

The Liouville-Arnold (LA)
theorem~\cite{arnold,
  fasano, jose, abraham1978foundations,
  Tanay:2020gfb}
  which states that, on a $2n$-dimensional symplectic manifold, if
  $\partial_t H=0$ and there are $n$ independent, 
  mutually commuting phase-space functions $F_{i}$,
   such that the level sets of these functions form compact
  and connected manifolds, then the system is integrable,
  and the above level sets are diffeomorphic to an $n$-torus.
  $H$ being one of these $F_i$'s implies that all the $F_i$'s
  are also constants since $\dot{F_i} = \{ F_i, H \} = 0$. 
  Hence we call these $F_i$'s the $n$ commuting constants.
  When $\Omega$ is exact, there is a globally well-defined potential
one-form $\Theta$ (such that $\Omega=d\Theta$), then in canonical
variables it will be
\begin{align}
  \Theta = \mathcal{P}_{i} d \mathcal{Q}^{i}      \label{potential}
  \,,
\end{align}
and the action variables can be computed via~\cite{arnold, fasano}
\begin{align}
  \mathcal{J}_{k}=\frac{1}{2 \pi} \oint_{\mathcal{C}_{k}} \Theta
  =\frac{1}{2 \pi} \oint_{\mathcal{C}_{k}}  \mathcal{P}_{i} d \mathcal{Q}^{i}
  \,,
  \label{eq:action_defined}
\end{align}
where $\mathcal{C}_{k}$ is any loop in the $k$th homotopy class on the
$n$-torus defined by the level sets
$F_i$=const. The above integral is insensitive to the choice of loop
in a certain homotopy class; see Proposition 11.2 of Ref.~\cite{fasano}.
However, the 2-sphere (and therefore our symplectic
manifold $B$) does not admit a global $\Theta$, as mentioned earlier.
  In such cases, the
actions are still well defined up to some global constants, but now
using integrals over areas instead of loops;
see Ref.~\cite{Tanay:2020gfb} for details.

Before ending this section, we briefly 
introduce the concept of Hamiltonian flows
and the associated Hamiltonian vector fields \cite{jose}. A  quantity
$f( \vec{\mathcal{P}}, \vec{\mathcal{Q}})$ defines a 
Hamiltonian vector field 
$\vec{X}_{f}$ via
$\pb{\cdot, f }=\pd/\pd \lambda$, such that it 
acts on another function
$g( \vec{\mathcal{P}}, \vec{\mathcal{Q}})$ as
$\pd g/\pd \lambda = \pb{g, f }$. The collection of the integral curves of
this vector field is referred to as 
the Hamiltonian flow of the field.

\section{The extended phase space: a tool to compute actions on spherical manifolds}      \label{sec:EPS}

In this section, there are instances where 
we first explain some subtle concepts informally,
before giving a more mathematically precise statement in the next paragraph.
The reader may choose to skip the more advanced wording at the 
expense of some depth.

\subsection{Motivation behind fictitious variables}           \label{sec:motivation}

In Ref.~\cite{Tanay:2020gfb}, we evaluated four
of the five action integrals for the 1.5PN BBH system.
The fifth action computation is a more complicated task and
this leads us to invent certain ``fictitious'', ``unmeasurable'' variables,
thereby extending the usual standard phase space (SPS) to the 
extended phase space (EPS). We now turn to explain the motivation behind them,
which has two facets.

Actions are well-defined on exact symplectic manifolds;
an exact symplectic manifold admits a global potential one-form
$\Theta = \vec{\mathcal{P}} \cdot d \vec{\mathcal{Q}}$).
While $\vec{R}$ and $\vec{P}$ live in
$T^{*}\RR^{3}$, which is exact (with $\Theta = \vec{P} \cdot d \vec{R}$), the same is not
true for the spin spherical symplectic manifolds,
thereby making the SPS non-exact; 
see Problem~2 of Homework~2 in Ref.~\cite{da2001lectures}.
Although the SPS is not exact, the EPS will be.
The two spaces will also be found to be equivalent 
(in a certain sense), which justifies the
computation of action in the EPS,
which we can then push forward to the SPS, since every
EPS point would map to an SPS point by construction.

The other more practical problem the EPS 
cures is that of computation of the action integral in
closed form.
In the SPS (with variables $R^i, P_i, \phi_a, S^z_a$ with $a= 1,2$), 
the  action integral is broken down into the orbital and spin sector contributions,
\begin{align}
\mathcal{J} &=\mathcal{J}^{\text {orb }}+\mathcal{J}^{\text {spin }}, \\
\mathcal{J}^{\text {orb }} & = \frac{1}{2 \pi} \oint_{\mathcal{C}_k}  P_{i} d R^{i},     \label{J5_orb_sector}    \\
\mathcal{J}_{A}^{\mathrm{spin}}  & =\frac{1}{2 \pi} \oint_{\mathcal{C}_k}  S_{A}^{z} d \phi_{A} \,.     \label{J5_spin_sector}
\end{align}
Now under the flow of $\Eff$,
the above orbital sector integral of 
Eq.~\eqref{J5_orb_sector} is easy to compute.
We state beforehand that the result comes out to be
\begin{align}
\mathcal{J}^{\text {orb }}=\frac{\left(S_{\mathrm{eff}} \cdot L\right) \Delta \lambda_{S_{\mathrm{eff}} \cdot L}}{ 2 \pi},
\end{align}
where $\Delta \lambda_{S_{\mathrm{eff}} \cdot L}$ is the flow amount under $\Eff$
(to be determined).
See Eqs.~\eqref{J5_int_a}-\eqref{J5_int_b} for the intermediate 
steps.\footnote{%
Eqs.~\eqref{J5_int_a}-\eqref{J5_int_b} are written for the EPS,
but if we neglect the spin sector terms (like $\mathcal{J}^{\text {spin }}$ 
and $P_{a i}~ {d R_{a}^{i}}/{d \lambda}$ with $a= 1, 2$), then these are also valid for the SPS.}
Now although the orbital sector of the action integral
 under the $\Eff$ flow is
easy to compute, we don't know how to compute the 
spin sector integral of Eq.~\eqref{J5_spin_sector}.
We again state beforehand that writing the orbital angular 
momentum $\vec{L}$ as a cross product of
a position $\vec{R}$ and a momentum $\vec{P}$ was critical to 
easily evaluating
$\mathcal{J}^{\text {orb }}$ under the $\Eff$ flow. 
This is something we can't do with 
the spin angular momenta $\vec{S}_a$
because $\vec{S}_a$ are considered to be fundamental
coordinates, not written as cross products of some
positions and momenta. 
As we will see, the EPS gets rid of all these problems, thus making
the action evaluation tractable.

\subsection{Introducing fictitious phase-space variables}           \label{sec:EPS_picture}

We refer to the phase space with coordinates 
$(\vec{R}, \vec{P}, \vec{S}_{1}, \vec{S}_{2})$ as the SPS,
the standard phase space. It is denoted by the letter $B$.
We now invent a new 18-dimensional extended phase space (EPS) $E=(T^{*}\RR^{3})^{3}$
with canonical coordinates $R^{i}, P_{i}, R_{a}^{i}, P_{ai}$ with
$a=1,2$, with canonical Poisson bracket algebra
\begin{align}
  \left\{R^{i}, P_{j}\right\}_{E}=\delta_{j}^{i} \,,
  \qquad
  \left\{ R_a^{i}, P_{bj}\right\}_{E}= \delta_{ab} \delta_{j}^{i} \,.
  \label{eq:PBextended}
\end{align}
Here we use the subscript $E$ to distinguish 
the Poisson brackets in
$E$ from those in $B$.
We call the $\vec{R}_{a}, \vec{P}_{a}$ variables the 
unmeasurable, fictitious variables. 
For contrast, we will sometimes refer to the SPS
coordinates $\vec{R}, \vec{P}, \vec{S}_{1}, \vec{S}_{2}$
as the observable coordinates.
We also demand that
 for an SPS point
$(\vec{R}, \vec{P}, \vec{S}_{1}, \vec{S}_{2} )$, the 
corresponding EPS point must satisfy
\begin{align}
  \vec{S}_1 &= \vec{R}_1 \times \vec{P}_1
  \,, &
  \vec{S}_2 &= \vec{R}_2 \times \vec{P}_2
  \,.
\end{align}
Of course, there are an
infinity of EPS points which correspond to the same SPS point.

In more advanced language,
this is a fiber bundle (with non-compact fibers) with
projection
\begin{align}         \label{projection}
  \pi: E\to B
  \,.
\end{align}
In coordinates, this projection takes a point in $E$ and sends it to
the point in $B$ where its $B$ coordinates are determined by%
\footnote{%
  A more sophisticated way to think of this projection is as follows.
  Think of the three-dimensional spin manifold, with coordinates
  $S_{a}^{i}$, as $\mathfrak{so}(3)^{*}$, the vector space dual to the
  Lie algebra $\mathfrak{so}(3)$ of the rotation group SO(3).  The
  dual of a Lie algebra naturally comes equipped with a Lie-Poisson
  structure (results developed by Kirillov, Kostant, and
  Souriau~\cite{MR1723696}).  The usual action of SO(3) on $\RR^{3}$
  induces a Hamiltonian action on its cotangent bundle $T^{*}\RR^{3}$
  (a Poisson manifold), analogous to our fiber coordinates
  $(R_{a}^{i},P_{aj})$.  From here we can build the dual map
  $T^{*}\RR^{3} \to \mathfrak{so}(3)^{*}$, which is the \emph{momentum
    map}~\cite{MR1723696}.  Our projection $\pi$ coincides with the
  momentum map.}
\begin{align}
  \label{eq:cross_prod}
  \pi( (\vec{R},\vec{P},
  \vec{R}_{1},\vec{P}_{1},
  \vec{R}_{2},\vec{P}_{2}
  ) )
  =
  (\vec{R},\vec{P},
  \vec{R}_{1}\times \vec{P}_{1},
  \vec{R}_{2}\times \vec{P}_{2})
  \,.
\end{align}
This is depicted in Fig.~\ref{fig:fiber}.
In this pictorial depiction of Fig.~\ref{fig:fiber}, a purely
vertical displacement in the EPS space
corresponds to changing the EPS coordinates in such a way that
the observable coordinates do not change.
To change the observable
coordinates, a horizontal motion is needed, both in the 
SPS and the EPS. In addition to Fig.~\ref{fig:fiber}, we also follow these simple
 pictorial conventions
in later figures like  Figs.~\ref{fig:eps_sps_1} and \ref{fig:eps_sps_2}.

\figC

Any function on $B$ can be pulled back with $\pi^{\star}$ to a function on
$E$. So we can evolve the  fictitious variables under the flow of the pulled-back
version of $H$.  While the fictitious variables can appear in intermediate calculations,
they are a mathematical convenience for the purpose of computing $\J$.
In the end, if physically observable quantities depend on 
$\vec{R}_{a}$ or $\vec{P}_{a}$, they must depend on them through
$\vec{S}_{a}$. In other words, the observable quantities must be functions
of only the observable coordinates.

In summary, this extended manifold $E$,
 the spins are now seen as cross products
of fictitious positions and momenta. Also, now $E$ is exact and admits a 
global potential one-form $\Theta_{E} = \vec{P} \cdot d \vec{R} +\vec{P}_1 \cdot d \vec{R}_1
+ \vec{P}_2 \cdot d \vec{R}_2 $. It is no longer a  product of 
a Cartesian manifold and two-spheres.
All three angular momenta
$\vec{L}, \vec{S}_1$ and $\vec{S}_2$ stand on equal
 mathematical footing.

\subsection{Comparing the EPS and SPS pictures}
\label{sec:comparing-eps-sps}

We can now sensibly talk about PBs on either the base SPS manifold $B$,
or the extended EPS manifold $E$, denoted as
$\pb{,}_{B}$ and $\pb{,}_{E}$, respectively. The former is computed using 
Eqs.~\eqref{PB1} and \eqref{PB2}, whereas the latter is computed using 
Eq.~\eqref{eq:PBextended}. Additional rules like Eqs.~\eqref{anti-symmetry}
and \eqref{chain-rule} apply universally to both $\pb{,}_{B}$ and $\pb{,}_{E}$.
Now that we have rid ourselves of the problematic features of the SPS,
the next natural question would be: are the two spaces (SPS and the EPS)
equivalent in some sense so as to justify action computation in the EPS,
instead of the SPS?
It is easy to check that, when acting on any two
functions $f$ and $g$ that only depend on the
SPS coordinates, the two PBs agree, since
Eqs.~\eqref{eq:PBextended} imply Eqs.~\eqref{PB1} and \eqref{PB2}.
Because of this crucial observation, we conclude that \textit{the SPS
picture and the EPS picture are equivalent} 
for the evolution of $f$
under the flow of $g$. In other words,
\begin{align}
  \label{eq:PB_preserved}
  \pi^{\star}\pb{f,g}_{B} = \pb{\pi^{\star}f,\pi^{\star}g}_{E}
 \, .
\end{align}
This means that either of the two pictures can be used to 
evolve the system under the $H$ flow.

We can state the above compatibility relation of the PBs in $B$ and $E$
in more advanced language of differential geometry.
Given some symplectic form $\Omega$,
its associated Poisson bracket $\pb{f,g}$ is found from
\begin{align}
  \pb{f,g}   =  \Omega^{-1}(df, dg)
  \,,
\end{align}
where $\Omega^{-1}$ is the bivector that is the inverse of $\Omega$,
$[\Omega^{-1}]^{ij}\Omega_{jk}=\delta^{i}_{k}$.
In our setting we have a symplectic form $\Omega_{B}$ in the SPS and
$\Omega_{E}$ in the EPS. Eq.~\eqref{eq:PB_preserved}, 
the compatibility condition between the two PBs can be re-expressed as
\begin{align}
  \label{eq:compat-geometric}
  \pi^{\star}  
  \left(
    \Omega_{B}^{-1} (df, dg)
  \right)  =  \Omega_{E}^{-1} ( \pi^{\star} df , \pi^{\star} dg )
  \,,
\end{align}
where $\pi^{\star}$ is the pullback induced by the projection map
$\pi$, and
$f,g:B\to\RR$.  Since the LHS is fiberwise constant,\footnote{
Fiberwise constancy means constancy when one moves along a fiber
through points which map to the same base point.
This means insensitivity to changing the fictitious variables (and
moving through the EPS), so long as
all these fictitious variables correspond to the same SPS point $(\vec{R}, \vec{P},
\vec{S}_1, \vec{S_2})$.   \label{fiberwise-constancy}} so is the RHS;
and so we can also consistently push forward this equality to $B$.
Since $f$ and $g$ are arbitrary, this compatibility and the 
definition of pushforward implies that
\begin{align}
  \label{eq:compat-geometric-push}
  \pi_{\star}(\Omega_{E}^{-1}) = \Omega_{B}^{-1}
  \,,
\end{align}
where $\pi_{\star}$ is the pushforward.

The equivalency needs to be pushed even to the integrability arena:
the 1.5PN BBH system being integrable or chaotic must not depend on 
whether we choose to work in the SPS picture or the EPS one.
Fortunately, the two pictures are also equivalent when we investigate the
integrability of the system, following the LA theorem.
In the base SPS manifold, we have the required
$10/2=5$ mutually commuting constants to establish integrability:
$H, J^2, J_z, L^2, S_{\text{eff} }\cdot L$.
In the EPS picture, we also have the requisite $18/2=9$ commuting
constants required for integrability. Those are the five
constants already listed above,
 plus $S_a^2$ and $R_a \cdot P_a $ 
for $a=1,2$.
These nine constants are to be 
viewed as functions of the EPS coordinates.
Because of the integrable nature of the system,
 there are five (nine) action variables in
the SPS (EPS), and similarly for the angle variables.

An interesting question arises. Imagine two points $P$ and $Q$ in the EPS
which have the same SPS coordinates $\vec{R}, \vec{P}, \vec{S}_{1}$ and $\vec{S}_{2}$,
but some different fictitious coordinates
(shown in Fig.~\ref{fig:eps_sps_1}). If we were to flow under 
$f (\vec{R}, \vec{P}, \vec{S}_{1} (\vec{R}_a,
 \vec{P}_a), \vec{S}_{2} (\vec{R}_a, \vec{P}_a) )$, 
 with $P$ and $Q$ as starting
points for a fixed amount $\lambda_0$, then are the SPS coordinates of the two final 
points the same? In other words, is
 $\pi( P'(P, f, \lambda_0) ) = \pi( Q'(Q, f, \lambda_0) )$?
The primes denote the final point reached at the end of the flow.
It is easy to check that the answer to the above question is 
`yes', and this is due to the compatibility of the PBs 
(Eq.~\eqref{eq:PB_preserved} or Eq.~\eqref{eq:compat-geometric-push}).
In other words, when flowing under  $f$
 in the EPS, the SPS coordinates of the final point reached by the flow
 depends only on $f$, $\lambda_0$ 
 and the SPS coordinates (and not the fictitious coordinates)
of the starting point.
This is not just a desirable but  a necessary
 feature because it assures us that among an infinity of EPS
 configurations  (lying within a single fiber)
  that are compatible with a given SPS configuration,
  we can choose to work with any one of them.

\figD

We can state the same result in the language
of Hamiltonian vector fields. We denote the Hamiltonian vector
field generated by the flow under $f$ (whether in the SPS or
in the EPS) with
\begin{align}
  X^{B}_{f} &\equiv \pb{\cdot, f}_{B} = \Omega_{B}^{-1}(\cdot, df)  ,   \\
  X^{E}_{\pi^{\star}f} &\equiv \pb{\cdot, \pi^{\star}f}_{E} = \Omega_{E}^{-1}(\cdot, d(\pi^{\star}f))
  = \Omega_{E}^{-1}(\cdot, \pi^{\star}(df))    ,
\end{align}
where the Poisson bracket on the EPS $\pb{\cdot,\cdot}_{E}$ acts on
the pullback $\pi^{\star}f$ of the function
$f = f (\vec{R}, \vec{P}, \vec{S}_{1} , \vec{S}_{2} )$.
Now the compatibility of the
brackets in the SPS and the EPS (Eqs.~\eqref{eq:PB_preserved}
and \eqref{eq:compat-geometric-push}) tells us that
\begin{align}
  \pi_{\star} (X^{E}_{(\pi^{\star}f)}) = X^{B}_{f}, 
\end{align}
i.e., the SPS vector field is the pushforward of the EPS
vector field. This is equivalent to the result arrived at in the previous
paragraph.

\subsection{Strategy to compute the action}                 \label{strategy}

Since the EPS and SPS are equivalent when acting on SPS functions,
we can use either of them for our calculations.
As already remarked in Sec.~\ref{sec:motivation}, we
don't know how to compute the fifth action in the SPS.
So we now turn to computing the fifth action in the EPS via
\begin{align}
\mathcal{J}_k   =  \frac{1}{2 \pi} \oint_{\mathcal{C}_{k}}   \left(    \vec{P} \cdot d \vec{R}   +     \vec{P}_1 \cdot d \vec{R}_1   +     \vec{P}_2 \cdot d \vec{R}_2  \right)  ,  \label{action-EPS}
\end{align}
which interestingly is tractable.
We state in advance the necessary result that the fifth action
(Eq.~\eqref{eq:action_B}) in the EPS is
fiberwise constant (see Footnote \ref{fiberwise-constancy}),
 meaning it can be written in terms of only the observable
coordinates ($\vec{R}, \vec{P}, \vec{S}_1, \vec{S}_2$).
In other words, the dependence of this action
on the unmeasurable variables occurs only through the combinations
$\vec{S}_1 = \vec{R}_{1}\times \vec{P}_{1}$ and $\vec{S}_2 = \vec{R}_{2}\times \vec{P}_{2}$.
This makes it possible to treat the fifth action as a function of only the
SPS coordinates.

Another important question arises.  Since we are computing the fifth
action in the EPS, do we have a legitimate action in the SPS?
The answer is `yes' in a certain sense (as explained below),
although due to the SPS-EPS equivalency, we can in principle, 
totally disregard the  SPS and work only in the EPS, through and through.
Eq.~\eqref{eq:action_defined} is the popular loop-integral
definition of action.
This action has an important property that under its flow by 
$2 \pi$ (and not any smaller amount), we get a closed 
loop.\footnote{See the proof of Theorem 11.6 of
Ref.~\cite{fasano} to arrive at this conclusion.}
In fact, this is such an important feature that it can also serve as
another definition of the action (call it the ``loop-flow definition'').
We will use the loop-integral definition to compute the action in the EPS,
 and we show in Appendix~\ref{SPS-action} that the 
 pushforward of this action to SPS
 satisfies the loop-flow definition of action.

We make a few closing remarks before we turn
 to the evaluation of the fifth action integral of Eq.~\eqref{action-EPS}.
We have numerically verified that
flowing by $2 \pi$ under the fifth action
 (to-be-computed from Eq.~\eqref{action-EPS})
yields a closed loop (as required by the loop-flow definition), 
within numerical errors,
whether the action is treated as an SPS function or an EPS function.
Although, the first four action integrals
computed in Ref.~\cite{Tanay:2020gfb} were done in the SPS,
we could have also computed them in the EPS, and then pushforwarded
these integrals to the SPS. The results would be the same as the 
four action integrals already presented in 
Ref.~\cite{Tanay:2020gfb}, except for
some irrelevant additive constants.
In summary, the equivalence of the two pictures
(in terms of integrability, action-angle variables, and
most importantly, the evolution under a flow associated with any observable),
the global exactness of the symplectic form $\Omega_{E}$, and the ease
of evaluation of the action variables, make us prefer the EPS over the SPS
 for the  action computation.

\section{Computing the fifth action}
\label{sec:comp-fifth-action}

Four out of the five actions were already presented in Ref.~\cite{Tanay:2020gfb}.
 Here we compute the fifth one.
For the fifth action, we generate a closed loop on the invariant
$n$-torus by flowing under $\Eff$, and other commuting constants.
 After flowing under $\Eff$ by a 
 certain amount $\Delta \lambda_{\Eff}$ (to be computed), although the
mutual angles between $(\vec{L},\vec{S}_{1},\vec{S}_{2})$ return to
their original values, these individual vectors have not. So we have
not formed a closed loop yet. However, additional flows under
$J^{2},L^{2},S_{1}^{2}$, and $S_{2}^{2}$ will close the loop 
(shown in Appendix \ref{flow_under_SeffdL}),
and at the same time ensuring that this loop  is in a different
homotopy class than the four associated to the other actions. We will see 
that we do not need to flow along 
$H$ or $J_{z}$ for the fifth action computation.
The fifth action integral can be computed piecewise as five integrals,
\begin{align}
{\mathcal{J}_5}  =   {\mathcal{J}}_{\Eff}   + {\mathcal{J}}_{J^2} + {\mathcal{J}}_{L^2} + {\mathcal{J}}_{S_1^2}  + {\mathcal{J}}_{S_2^2} ,    \label{eq:action_decomposed}
\end{align}
where each part corresponds to the segment generated by flowing under
the quantity in the subscript.

Focusing on ${\mathcal{J}}_{\Eff}$, we will need the evolution equations
under the flow of $\Eff$ in the EPS, which read
\begin{subequations}
\begin{align}
\frac{d \vec{R}}{d \lambda}            &=   \vec{S}_{\eff} \times \vec{R}   , \\
\frac{d \vec{P}}{d \lambda}            & =   \vec{S}_{\eff}   \times \vec{P}  ,  \\
\frac{d \vec{R}_{a}}{d \lambda}  &= \sigma_{a}\left(\vec{L} \times \vec{R}_{a}\right) ,\\ 
\frac{d \vec{P}_{a}}{d \lambda}  &= \sigma_{a}\left(\vec{L} \times \vec{P}_{a}\right),
\end{align}
\end{subequations}
and they imply
\begin{subequations}
\begin{align}
\frac{d \vec{L}}{d \lambda}            &=   \vec{S}_{\eff} \times \vec{L}   , \\
\frac{d \vec{S}_{a}}{d \lambda}  &= \sigma_{a}\left(\vec{L} \times \vec{S}_{a}\right) .  
\end{align}
\end{subequations}
From these evolution equations we have
\begin{align}
  &   2 \pi {\mathcal{J}}_{\Eff}   = 2 \pi ( \mathcal{J}^{\text{orb}}  + \mathcal{J}^{\text{spin}} )       \label{J5_int_a}     \\  
    &=     \int_{\lambda_i}^{\lambda_f}  \left(  P_i \frac{d R^i}{d \lambda}    +  P_{1i} \frac{d R_1^i}{d \lambda}    +  P_{2i} \frac{d R_2^i}{d \lambda}    \right)  d \lambda      \nonumber \\
   & =      \int_{\lambda_i}^{\lambda_f}   \left(   \vec{P} \cdot ( \vec{S}_{\eff}  \times \vec{R} )  +     \vec{P}_{1} \cdot (  \sigma_{1} \vec{L}  \times \vec{R}_1 )    \right.   \nonumber  \\ 
   & ~~~~~~~    \left. +     \vec{P}_{2} \cdot (  \sigma_{2} \vec{L}  \times \vec{R}_2 )  \right)     d \lambda     \\
   & =   2   \int_{\lambda_i}^{\lambda_f}   \left(   {S}_{\eff}  \cdot {L}    \right)     d \lambda    =    2     ({S}_{\eff}  \cdot {L} )  \Delta \lambda_{\Eff} ,   \\
 &    {\mathcal{J}}_{\Eff}    =  \frac{     ({S}_{\eff}  \cdot {L} )  \Delta \lambda_{\Eff}}{ \pi}          \label{J5_int_b} 
\,,
\end{align}  
with $\Delta \lambda_{\Eff} = \lambda_{f}-\lambda_{i}$ being the
required flow parameter amount (for the mutual angles of $(\vec{L},\vec{S}_{1},\vec{S}_{2})$ to be restored).
We could pull $\Eff$ out of the integral since it is a constant under the flow of $\Eff$. 
After performing similar calculations, we can also show that (see also Sec. III-A of Ref.~\cite{Tanay:2020gfb})
\begin{subequations}
\begin{align}
  \mathcal{J}_{J^2}  = \frac{J^2 \Delta \lambda_{J^2} }{\pi}       ,  \\
  \mathcal{J}_{L^2}  = \frac{L^2 \Delta \lambda_{L^2}}{\pi}       ,  \\
  \mathcal{J}_{S_1^2}  = \frac{S_1^2  \Delta \lambda_{S_1^2} }{\pi}   ,  \\
  \mathcal{J}_{S_2^2}  = \frac{S_2^2 \Delta \lambda_{S_2^2}}{\pi}     ,
\end{align}
\end{subequations}
where the quantities $\Delta \lambda_i$'s
are the flow amounts required to close the loop
 under the corresponding commuting constant in 
 the subscript. This finally renders the fifth action to be
\begin{align}
\mathcal{J}_5   &  = \frac{1}{\pi}  \bigg\{   ({S}_{\eff}  \cdot {L} )  \Delta \lambda_{\Eff} + J^2 \Delta \lambda_{J^2} + L^2 \Delta \lambda_{L^2}  \nonumber \\ 
                          &  ~~~~~  +  S_1^2  \Delta \lambda_{S_1^2} + S_2^2 \Delta \lambda_{S_2^2}   \bigg\},            \label{eq:action_B}
\end{align}
which means that the fifth action computation has now
boiled down to computing the five flow amounts $\Delta \lambda_i$'s.
Due to the tedious nature of the computation of these five parameter 
flow amounts, they have been relegated to Appendix~\ref{flow_under_SeffdL}.

Summarizing, the fifth action is given by Eq.~(\ref{eq:action_B}), 
where the $\Delta \lambda$'s are
presented in Eqs.~(\ref{lambda_1}), (\ref{lambda_2}), (\ref{lambda_3}),
(\ref{lambda_4}) and (\ref{lambda_5}).
Because
our derivation assumed $m_1 > m_2$, this expression of the action
is not manifestly symmetric under the
exchange $1 \leftrightarrow 2$ (labels of the two black holes).
However, as discussed in the text after Eq.~\eqref{mass_case}, the
symmetry can be restored simply.
Note that in Ref.~\cite{Tanay:2020gfb},
the flows under $J^{2},J_{z}$, and $L^{2}$ individually form
closed loops. This implies that the associated actions are
functions of just these individual conserved quantities.
Meanwhile, to get a closed loop for the fifth action evaluation,
we need to flow under all of $J, L, \Eff, S_{1}$, and $S_{2}$, 
which makes the fifth action a function of all these five quantities.

\subsection*{Fifth action in the equal mass case}

The above result for the fifth action in Eq.~\eqref{eq:action_B} is not
manifestly finite in the equal mass limit: there are many factors of
$(\sigma_{1}-\sigma_{2})$ which vanish in this limit, including some in
denominators. We have checked numerically that the equal mass limit of
$\mathcal{J}_5$ is finite, but trying to take this limit analytically
is cumbersome.  There is however a simpler way, and the solvability of
the equal-mass case has been independently investigated in the
literature, albeit in the orbit- and precession-averaged
approach~\cite{Gerosa:2016aus}.

Working with only the SPS variables,
when $\sigma_1 = \sigma_2$ (equal-mass case), 
it is easy to check that $\vec{S}_1 \cdot \vec{S}_2$,
 along with $H, J^2, L^2$, and $J_z$
forms a set of five mutually commuting constants. In fact, $\Eff$ can then be seen as a function of these five constants, and
is therefore no longer an independent constant. 
It can be checked that under the flow of $\vec{S}_1 \cdot \vec{S}_2$
we have the flow equations (with $\vec{S} \equiv \vec{S}_1 + \vec{S}_2$)
\begin{align}
\pb{\vec{S}_1 , \vec{S}_1 \cdot \vec{S}_2} =   \vec{S} \times \vec{S}_1 =  \pb{ \vec{S}_1 \cdot \vec{S}_2 , \vec{S}_2 }   = \vec{S}_2  \times  \vec{S} ,
\end{align}
which imply that both the spin vectors 
rotate around $\vec{S}$,
 which itself remains fixed under this flow.
 $\vec{R}$ and $\vec{P}$ don't move and hence only the 
 spin sectors contribute to the action integral.
At this point, we can simply use the result
 of Eq.~(28) of Ref.~\cite{Tanay:2020gfb} with $\hat{n} = \vec{S}/S$,
which gives our fifth action variable for the equal mass case as
\begin{align}
\widetilde{\mathcal{J}}_{5(m_1 = m_2)}   = (\vec{S}_1 + \vec{S}_2) \cdot \vec{S}/S = S   .
\end{align}
The reason we used a tilde in the above equation is because $\widetilde{\mathcal{J}}_{5(m_1 = m_2)} $
need not be the equal mass limit of $\mathcal{J}_5$, since
action variables of a system are not unique; see Proposition 11.3 of Ref.~\cite{fasano}.

Finally, using the equal mass relations
\begin{align}
J^2    & = L^2 + S_1^2 +S_2^2 + 2 ( \vec{L} \cdot \vec{S} + \vec{S}_1 \cdot \vec{S}_2 )  , \\
\Eff     & =  \frac{7}{4}  \vec{L} \cdot \vec{S}      , \\
S^2   &   = S_1^2 + S_2^2 + 2 \vec{S}_1 \cdot \vec{S}_2,
\end{align}
in Eq.~(38) of Ref.~\cite{Tanay:2020gfb}, it is possible to 
arrive at an equation connecting the Hamiltonian with the actions.
Performing a PN series inversion thereafter, one can write
an explicit expression for the Hamiltonian in terms of the actions, up to
1.5PN. This can be used to explicitly obtain the frequencies of the
system via $\omega^{i} = \partial H / \partial \mathcal{J}_i$ for the
equal-mass case.

\section{Fifth action at the leading PN order}          \label{sec:lin_action}

The action variable given by Eq.~\eqref{eq:action_B} is in exact form with respect to the 1.5PN Hamiltonian $H$. It is a worthwhile exercise 
to write the leading order contribution of this action because it is a much shorter expression than the ``exact'' one. This is
in the same spirit as the expression of the fourth action variable
as  a PN series which was presented in Eq.~(38)
 of Ref.~\cite{Tanay:2020gfb}.
Another advantage is that we can then write $\Eff$
 in terms of the actions, including the fifth one 
 (discussed below), which when used with
Eq.~(38) of Ref.~\cite{Tanay:2020gfb} 
can give an expression for Hamiltonian in terms of the actions.

Note that out of the five actions: 
$J, L, J_z, \mathcal{J}_4$, and $\mathcal{J}_5$ 
(see Ref.~\cite{Tanay:2020gfb}
for the first four), the first two coincide with each other
 at 1PN order due to the absence of spins. 
 The next important action variable at 1PN 
is the 1PN version of $\mathcal{J}_4$~\cite{damour1988}.
$J_z$ is irrelevant when it comes to computing frequencies
since the Hamiltonian is never  a function of $J_z$.
 This explains the presence of only two frequencies 
(resulting from effectively two actions) at 1PN. Now since $\mathcal{J}_5$ 
comes into play for the first time only at 
the 1.5PN order, it makes sense to expand it in a PN series
and work with the leading order term only, if we are working at 1.5PN.
We now turn our attention to extracting this leading order term.

We sketch the plan for how to obtain the 
leading PN contribution to $\mathcal{J}_5$. 
It comprises a couple of steps
which were performed in \textsc{Mathematica}.

\paragraph*{Step 1:} To start with, instead of writing the various quantities which make up $\mathcal{J}_5$
in terms of the five commuting constants, write them
only in terms of $\vec{L}, \vec{S}_1, \vec{S}_2, \sigma_1$ and $\sigma_2$ with the understanding that $\vec{S}_1$
and $\vec{S}_2$ are 0.5PN order higher than $\vec{L}$; 
see Ref.~\cite{Tanay:2020gfb} for more details on this.
Attach a formal PN order counting parameter
 $\epsilon$ to $\vec{S}_1$ and $\vec{S}_2$. This $\epsilon$ will
be used as a PN perturbative expansion parameter: 
every power of $\epsilon$ stands for an extra 0.5PN order.
At the end of the calculation, $\epsilon$ will be set equal to 1.
Writing various quantities of interest in terms of $\vec{L}, \vec{S}_1$, and $\vec{S}_2$ is imperative since it serves to expose the
PN powers explicitly. For example, $J^2 - L^2 =
\mathcal{O}(\epsilon^{1})$, though both $J^2$ and $L^2$ are
$\mathcal{O}(\epsilon^0)$. This becomes manifestly clear when
$J^2 - L^2$ is written in the above way.

\paragraph*{Step 2:} Instead of trying to series expand $\mathcal{J}_5$ directly in terms of $\epsilon$ in one go, we
first series expand various quantities that make up $\mathcal{J}_5$, and then use these expanded versions to
finally build up the series-expanded version of $\mathcal{J}_5$. As a first step, series expand the roots of the cubic expression of 
Eq.~\eqref{eq:cubic-1}, keeping terms up to 
${\cal{O}}(\epsilon^2)$.

\paragraph*{Step 3:} Series expand various other quantities that make up $\mathcal{J}_5$,
such as $k^2, B_1, B_2, D_1, D_2, \alpha_1$ and $\alpha_2$ in $\epsilon$ such that
the resulting expansions have two non-zero post-Newtonian terms. We don't have to
worry about series expanding
certain other quantities which make up $\Delta \lambda_4$ and $\Delta \lambda_5$, since 
they don't contribute to the fifth action variable at the leading
order.

\paragraph*{Step 4:} Using these series-expanded ingredients, build up $\mathcal{J}_5$ of
Eq.~(\ref{eq:action_B}). The PN orders of the five summands of $\mathcal{J}_5$
(as shown in Eq.~(\ref{eq:action_decomposed})) are schematically shown here as
\begin{align}
\mathcal{J}_{\Eff}  & =  \mathcal{O}(\epsilon)      ,      \\
\mathcal{J}_{J^2}  & =  {  \mathcal{J}_0}  \epsilon^0 +  \mathcal{O}(\epsilon)      ,      \\
\mathcal{J}_{L^2}  & =  {- \mathcal{J}_0} \epsilon^0 +  \mathcal{O}(\epsilon)      ,      \\
\mathcal{J}_{S_1^2}  & =    \mathcal{O}(\epsilon^2)      ,      \\
\mathcal{J}_{S_2^2}  & =    \mathcal{O}(\epsilon^2)      ,
\end{align}
where we have indicated that the leading order components of $\mathcal{J}_{J^2}$ 
and $\mathcal{J}_{L^2} $ cancel each other. Our leading order $\mathcal{J}_5$ is thus
the sum of the first three contributions. The last two contributions being at sub-leading orders can be dropped.
At this point we can set $\epsilon = 1$.

\paragraph*{Step 5:} At this point the resulting perturbative $\mathcal{J}_5$
is a function of $\vec{L}, \vec{S}_1,\vec{S}_2, \sigma_1, \sigma_2$
and dot products formed out of them. We still want to write this as
a function of the commuting constants only, keeping in line with the tradition followed in the
action-angle variables formalism. To do so, we eliminate $\vec{L} \cdot \vec{S}_1$ and $\vec{L} \cdot \vec{S}_2$
using the following results valid up to the leading PN order
\begin{subequations}                     \label{aux-100}
\begin{align}
\vec{L} \cdot \vec{S}_1  & \sim   \frac{2 \Eff   -  (J^2 - L^2  -  S_1^2  - S_2^2) \sigma_2}{2(\sigma_1 - \sigma_2)}    ,   \\
\vec{L} \cdot \vec{S}_2  & \sim -  \frac{2 \Eff  -  (J^2 - L^2  - S_1^2  - S_2^2)\sigma_1}{2(\sigma_1 - \sigma_2)}     .
\end{align}
\end{subequations}

With this and the following new definitions,
\begin{widetext}
\begin{align}
    \mathcal{A} &= -J^2 + L^2 + S_1^2 + S_2^2 , \\
    \mathcal{B}_1 &= 2 S_{\text{eff}} \cdot L + \mathcal{A} \sigma_1, \\
    \mathcal{B}_2 &= 2 S_{\text{eff}} \cdot L + \mathcal{A} \sigma_2, \\
    \mathcal{C}_1 &= \sqrt{L^2 S_2^2 - \frac{\mathcal{B}_1^2}{4 (\sigma_1 - \sigma_2)^2}} ,\\
    \mathcal{C}_2 &= \sqrt{L^2 S_1^2 - \frac{\mathcal{B}_2^2}{4 (\sigma_1 - \sigma_2)^2}}, \\
    \mathcal{D}  & = 
\sqrt{-\left( \frac{-\mathcal{B}_1^2 - \mathcal{B}_2^2 - 4 (2 \mathcal{C}_1 \mathcal{C}_2 - L^2 (S_1^2 + S_2^2)) (\sigma_1 - \sigma_2)^2}{\mathcal{B}_1^2 + \mathcal{B}_2^2 - 4 (2 \mathcal{C}_1 \mathcal{C}_2 + L^2 (S_1^2 + S_2^2)) (\sigma_1 - \sigma_2)^2} \right)},
\end{align}
the corrected expression of the fifth action 
variable at the leading PN order becomes 
\begin{equation}
\begin{split}
    \mathcal{J}_5  & \sim  \frac{1}{4 L (\sigma_1 - \sigma_2) \sqrt{ 1 + \frac{4 \mathcal{C}_1 \mathcal{C}_2}{-2 \mathcal{C}_1 \mathcal{C}_2 - L^2 (S_1^2 + S_2^2) + \frac{\mathcal{B}_1^2 + \mathcal{B}_2^2}{4 (\sigma_1 - \sigma_2)^2}} } \left(\mathcal{B}_1^2 + \mathcal{B}_2^2 - 4 (2 \mathcal{C}_1 \mathcal{C}_2 + L^2 (S_1^2 + S_2^2)) (\sigma_1 - \sigma_2)^2\right)} \\
    &\quad \times \Bigg[ -\mathcal{B}_1^3 (\mathcal{D} - 1) - \mathcal{B}_2^3 (\mathcal{D} - 1) - \mathcal{B}_1^2 \mathcal{B}_2 (\mathcal{D} + 1) - \mathcal{B}_1 \mathcal{B}_2^2 (\mathcal{D} + 1) \\
    &\quad + 4 (\sigma_1 - \sigma_2)^2 \Big( \mathcal{B}_1 (2 \mathcal{C}_1 \mathcal{C}_2 \mathcal{D} + L^2 ((\mathcal{D} + 1) S_1^2 + (\mathcal{D} - 1) S_2^2)) + \mathcal{B}_2 (2 \mathcal{C}_1 \mathcal{C}_2 \mathcal{D} + L^2 ((\mathcal{D} - 1) S_1^2 + (\mathcal{D} + 1) S_2^2)) \Big) \Bigg].
\end{split}          \label{lin-5th-action}
\end{equation}
\end{widetext}

We could have chosen to eliminate $\vec{L} \cdot \vec{S}_1$ and $\vec{L} \cdot \vec{S}_2$
using slightly modified forms of Eqs.~\eqref{aux-100} by simply ignoring $S_1^2$ and $S_2^2$ terms in
the numerator. These modified forms of Eqs.~\eqref{aux-100} and the resulting modified form of the leading order contribution
to the fifth action would still agree with the original results (Eqs.~\eqref{aux-100} and Eq.~\eqref{lin-5th-action}) up to
the leading PN order. The above expression of linearized fifth action
is not manifestly symmetric with respect to the label exchange $1
\leftrightarrow 2$.
This is because from the beginning, we assumed $m_1 > m_2$
while deriving the 1.5PN exact fifth action; 
see the text after Eq.~\eqref{mass_case}.
We can easily make this leading PN order
 version of fifth action symmetric by replacing
$(\sigma_1  - \sigma_2)$ with 
$-|\sigma_1  - \sigma_2|$  in the
denominator of the RHS of Eq.~\eqref{lin-5th-action}.
This is because
$m_1 > m_2 \implies \sigma_1 < \sigma_2 $.
We have  numerically verified that $\mathcal{J}_5$ as 
presented in Eq.~\eqref{lin-5th-action}
above converges to the exact 1.5PN version in the 
limit of small PN parameter ($S_1, S_2 \ll  L$ ).

\section{Frequencies and angle variables}     \label{sec:freq}

\subsection{Computing the frequencies}
\label{sec:frequencies}

Since we have an integrable Hamiltonian system,
the Hamiltonian is a function of the actions and not the angles,
 though it may not
be possible to write $H$ explicitly in terms of the actions.
In terms of the actions, the equations of motion for the
respective angle variables are trivial,
\begin{align}
  \dot{\theta}^{i} = \frac{\pd H}{\pd \mathcal{J}_{i}} = \omega^{i}(\vec{\mathcal{J}})
  \,.
\end{align}
As a consequence, the usual phase space variables are all
multiply-periodic functions of all of the angle
variables. Concretely, this means a
Fourier transform of some regular coordinate would consist of a forest
of delta function peaks at integer-linear combinations of the fundamental
frequencies $\omega^{i}$~\cite{landau1976mechanics}.
Additionally, if we know the frequencies, we can locate resonances --- 
where the ratio of two frequencies is a rational number --- which are
key to the KAM theorem and the onset of chaos.

With $\vec{C}$ standing for the vector of all five mutually commuting constants,
$H$ being one of these $C_i$'s, $H$ is automatically a function of $\vec{C}$.
In principle, once can invert $\vec{\mathcal{J}}(\vec{C})$ 
(at least locally, via the inverse
function theorem) for $\vec{\mathcal{C}}(\vec{\mathcal{J}})$, and thus
find an explicit expression for $H(\vec{\mathcal{J}})$ paving the road for the 
computation of the frequencies $\omega^{i}$'s.
But this is not necessary.

Instead, we follow the approach given in Appendix~A
of Ref.~\cite{Schmidt:2002qk} to find the frequencies as functions of the
constants of motion, via the Jacobian matrix between the five
$C_i$'s and the five $\mathcal{J}_i$'s.  
For the purpose of frequency computations, we take our $C_i$'s to be
(in this specific order)
$\vec{\mathcal{C}} =
\left\{
  J , J_{z}, L , H, \Eff
\right\}$.  As two of us showed in Ref.~\cite{Tanay:2020gfb},
the first three of these are already action variables.
We take the order of the actions to be $\vec{\mathcal{J}} =
\left\{
  J , J_{z}, L, \mathcal{J}_{4}, \mathcal{J}_{5}
\right\}$.
The expression for $\mathcal{J}_{4}$ was given as an explicit function
of $(H,L,\Eff)$ in Ref.~\cite{Tanay:2020gfb}.
The Jacobian matrix $\pd
\mathcal{J}^{i}/\pd{C}^{j}$ can be found explicitly, since we
have analytical expressions for
$\vec{\mathcal{J}}(\vec{{C}})$.  This matrix is somewhat
sparse, given by
\begin{align}
  \label{eq:dJdC-matrix}
  \frac{\pd \mathcal{J}^{i}}{\pd {C}^{j}} =
  \begin{bmatrix}
    1 & 0 & 0 & 0 & 0 \\
    0 & 1 & 0 & 0 & 0 \\
    0 & 0 & 1 & 0 & 0 \\
    0 & 0 & \frac{\pd \mathcal{J}_{4}}{\pd L} & \frac{\pd \mathcal{J}_{4}}{\pd H} &  \frac{\pd \mathcal{J}_{4}}{\pd (\Eff)} \\
    \frac{\pd \mathcal{J}_{5}}{\pd J} & 0 & \frac{\pd \mathcal{J}_{5}}{\pd L} & 0 &  \frac{\pd \mathcal{J}_{5}}{\pd (\Eff)}
  \end{bmatrix}
  \,.
\end{align}
Now we use the simple fact that the Jacobian $\pd {C}^{i}/\pd
\mathcal{J}^{j}$ is the inverse of this matrix (assuming it is full rank),
\begin{align}
  \frac{\pd \mathcal{J}^{i}}{\pd {C}^{j}} \frac{\pd {C}^{j}}{\pd \mathcal{J}^{k}} &= \delta^{i}{}_{k} \,,\\
  \frac{\pd {\vec{C}}}{\pd \vec{\mathcal{J}}} &=
  \left[
    \frac{\pd \vec{\mathcal{J}}}{\pd \vec{C}}
  \right]^{-1}
  \,.
\end{align}
Because of the sparsity of the matrix in Eq.~\eqref{eq:dJdC-matrix}, we
directly invert and find the only nonvanishing coefficients in the
inverse are
\begin{align}
  \label{eq:dCdJ-matrix}
  \frac{\pd \mathcal{C}^{i}}{\pd \mathcal{J}^{j}} =
  \begin{bmatrix}
    1 & 0 & 0 & 0 & 0 \\
    0 & 1 & 0 & 0 & 0 \\
    0 & 0 & 1 & 0 & 0 \\
    \frac{\pd H}{\pd J} & 0 & \frac{\pd H}{\pd L} & \frac{\pd H}{\pd \mathcal{J}_{4}} &  \frac{\pd H}{\pd \mathcal{J}_{5}} \\
    \frac{\pd (\Eff)}{\pd J} & 0 & \frac{\pd (\Eff)}{\pd L} & 0 &  \frac{\pd (\Eff)}{\pd \mathcal{J}_{5}}
  \end{bmatrix}
  \,.
\end{align}
The frequencies we seek are in the fourth row of this matrix.
Matrix inversion yields the following expressions for the frequencies:
\begin{subequations}              \label{freq_eqn}
\begin{align}
  \label{eq:dCdJ-entries}
  \frac{\pd H}{\pd J} = \omega^{1} &= \frac{(\pd \mathcal{J}_4/\pd (\Eff)) (\pd \mathcal{J}_5/\pd J)}{(\pd \mathcal{J}_4/\pd H) (\pd \mathcal{J}_5/\pd (\Eff))} \,,\\
              \frac{  \pd H}{\pd J_{z}}    = \omega^{2} &  = 0  , \\
  \frac{\pd H}{\pd L} = \omega^{3} &= \Big[(\pd \mathcal{J}_4/\pd (\Eff)) (\pd \mathcal{J}_5/\pd L) \\
  & \quad - (\pd \mathcal{J}_4/\pd L) (\pd \mathcal{J}_5/\pd (\Eff))\Big] \nn \\
  & \quad \times (\pd \mathcal{J}_4/\pd H)^{-1} (\pd \mathcal{J}_5/\pd (\Eff))^{-1}
  \,, \nn \\
  \frac{\pd H}{\pd \mathcal{J}_{4}} = \omega^{4} &= (\pd \mathcal{J}_{4}/\pd H)^{-1} \,,\\
  \frac{\pd H}{\pd \mathcal{J}_{5}} = \omega^{5} &= -\frac{\pd \mathcal{J}_{4}/\pd (\Eff)}{(\pd \mathcal{J}_{4}/\pd H)(\pd \mathcal{J}_{5}/\pd (\Eff))}   .
\end{align}
\end{subequations}
The frequency $\omega^{2} = \pd H/\pd J_{z}$ vanishes since $H$
cannot depend on $J_{z}$, to preserve SO(3) symmetry.
The derivatives of $\mathcal{J}_{4}$ with respect to $(H,L,\Eff)$ are
easy to compute from the explicit expression given in Eq.~(38) of Ref.~\cite{Tanay:2020gfb}.
Taking the derivatives of $\mathcal{J}_5$ in Eqs.~\eqref{freq_eqn}
involves many intermediate quantities that arise from the chain rule,
and are presented in Appendix~\ref{derivative calculations}.

\subsection{The angle variables}     \label{angle-subsec}

Canonical perturbation theory~\cite{goldstein2013classical, jose}
 has the potential to furnish 2PN action-angle 
variables when supplied with 1.5PN ones.
To use this tool, we want to be able to
express perturbations to the Hamiltonian (namely, higher PN order
terms) as functions of the angle variables which are canonically
conjugate to the actions.  One of these angles --- the mean anomaly,
which is conjugate to our $\mathcal{J}_{4}$ --- has been presented
previously in the literature, in pieces.
We have explicitly checked that the Poisson bracket between
$\mathcal{J}_4$ the 1.5PN mean anomaly (combining 1PN and 1.5PN pieces of
the results from Refs.~\cite{1985AIHS...43..107D} and
\cite{Gopakumar:2011zz}) is 1, up to 1.5PN 
order.\footnote{The result in Refs.~\cite{1985AIHS...43..107D} (Eq.~(7.1~a))
does not have the 1.5PN piece,
whereas the result in Ref.~\cite{Gopakumar:2011zz} 
(Eq.~(11b)) is missing the 1PN piece.}

\subsubsection{Constructing angle variables}

We now lay out a roadmap on how to implicitly construct the rest of
the angle variables on the invariant tori of constant
$\vec{\mathcal{J}}$ (or constant $\vec{C}$).
To be more precise, we show how to obtain the standard phase-space
coordinates $(\vec{\mathcal{P}} , \vec{\mathcal{Q}})$
as explicit functions of action-angle variables $(  \vec{\mathcal{J}}
, \vec{\theta} )$.  This is in fact the more useful transformation
(rather than $(  \vec{\mathcal{J}} , \vec{\theta} )$ as explicit
functions of $(\vec{\mathcal{P}} , \vec{\mathcal{Q}})$) for
canonical perturbation theory, since we will need to transform the 2PN
and higher Hamiltonian (which is
given as explicit function of 
$(\vec{\mathcal{P}} , \vec{\mathcal{Q}})$) into action-angle variables.

The method to assign angle variables on invariant tori
is straightforward. Pick a fiducial point $P_{0}$ on an invariant torus,
and give it angle
coordinates  $\vec{0}  \equiv (0,\ldots,0)$. 
Then every other point on this same
torus, with angle coordinates $\theta^{i}$, is reached by integrating
a flow from $P_{0}$ by  amounts $\theta^{i}$ under each of
the actions $\mathcal{J}_{i}$.
This is because the flow parameter is
in fact the angle parameter: $d \theta^i/d \lambda^{j}
=\pb{ \theta^i , \mathcal{J}_ j } = \delta^{i}_j$.
The Poisson brackets evaluating to Kronecker delta follows because
$ \theta^i$ and $\mathcal{J}_ j$ are canonically conjugate coordinates;
see Theorem~10.17 of Ref. \cite{fasano}.
Since the actions commute, we are free to flow under these
actions in any order.

The construction explained above was only on an individual torus.
The only requirement for extending these variables to being full phase
space variables is that the choice of fiducial point
$P_{0}(\vec{\mathcal{J}})$ is smooth in $\vec{\mathcal{J}}$.  Given
any choice of angle variables, we can always re-parameterize them by
adding a constant that is a smooth function of $\vec{\mathcal{J}}$.
That is, if $\theta^{i}$ are angle variables, then so are
$\bar{\theta}^{i} = \theta^{i} + \delta\theta^{i}(\vec{\mathcal{J}})$,
with smooth $\delta\theta^{i}$, which can be verified by taking
Poisson brackets: 
$\pb{ \bar{\theta}^i , \mathcal{J}_ j } = \delta^{i}_j$.  
Some of these angle variables may be simpler than
others, but here we are only interested in finding one such
construction.

So now, the problem of assigning the angle coordinates on the torus has
been transformed into that of flowing under all the actions, one 
by one, by amounts equal to the angle coordinates of the point
whose angles are desired (assuming that the starting point
had $\vec{\theta} = \vec{0}$).
To integrate the equations under the flow associated 
with any of the five actions, we start with
\begin{align}
\frac{d \xi }{ d \lambda}  & = \pb{ \xi , \mathcal{J}_i  (\vec{C})}     ,  \nonumber \\
                                   &  =  \pb{ \xi , C_j }  \frac{ \partial \mathcal{J}_i }{\partial  C_j} ,
\end{align}
where $\xi$ is any one of the phase space coordinates.
This is the same sparse matrix $\pd\mathcal{J}_{i}/\pd C_{j}$ which appeared
in the previous section in Eq.~\eqref{eq:dJdC-matrix}.
The matrix $ \partial \mathcal{J}_i/\partial  C_j$ is a function of
only the $\vec{C}$'s, and thus is constant on each torus and each of the
flows we consider.  Hence, integrating the above equation boils down to
integrating under the flow of the $C_i$'s.
We will now briefly explain how to
obtain the solution for the flow under 
each of the $C_i$'s one by one.

\subsubsection{Solutions to flow under the commuting constants}

The solution for flow under $H$ has been given in 
Ref.~\cite{Samanta:2022yfe}; it has been termed as the 
``standard solution'' there. It is found by filling in the gaps in the
solution provided in 
Ref.~\cite{cho2019analytic}.\footnote{Reference \cite{cho2019analytic}
ignored the 
1PN Hamiltonian throughout for brevity since the authors deemed it
straightforward. Eqs.~(3.28-c,~d) of this article have typos. \label{cho-footnote}}
The solution for the flow under $\Eff$ is constructed
in Appendix~\ref{flow_under_SeffdL}, with minor caveats.
Eqs.~\ref{mutual-angle-sol-2}, \ref{eq:overflow_phi_L}, and
 \ref{eq:Delta_phi_soln} in Appendix~\ref{flow_under_SeffdL}
  collectively give solutions for $\vec{L}$
 and $\vec{R}$, but the appendix
 does not give explicit solutions for
 $\vec{P},~\vec{S}_1$ and $\vec{S}_2$. However, this is
 not a major hurdle for the following reasons. 
  The solution for
$\vec{P}$ can be easily found from that for
 $\vec{R}$ by noting that $P$ and
the angular offset between $\vec{R}$ and $\vec{P}$
 remain constant under the $\Eff$ flow.
Also, the solutions for $\vec{S}_1$ can 
be had using similar calculations as for
the solution for $\vec{L}$. Once we have $\vec{S}_1$,
 $\vec{S}_2$ can be found from
 $\vec{S}_2 = \vec{J} - \vec{L} - \vec{S}_1$, and the fact that
 $\vec{J}$ does not change under the $\Eff$ flow.

It now remains to
show how to integrate under the flow of the 
remaining three $C_i$'s, ($J^2, J_z$ and $L^2$).
Sec.~III (specifically Eqs.~(21)-(23)) of Ref.~\cite{Tanay:2020gfb}
showed that
the equations for a flow under any of these quantities can be concisely written in a generalized form as
\begin{align}
\frac{ d \vec{V}}{d \lambda }   = \pb{\vec{V} , \mathcal{J}_i}  =   \vec{U} \times  \vec{V}.     \label{flow_rotation}
\end{align}
Here
$\vec{U}$ is the constant vector 
(under the respective flow) $2 \vec{J}, {\hat{z}}$, or $2\vec{L}$ 
when $C_i$ is
$J^2, J_z$, or $L^2$, respectively.
In the above equation, $\vec{V}$ stands for any
of $\vec{R}, \vec{P}, \vec{S}_1$, and $\vec{S}_2$, with the exception that
under the flow of $L^2$, spin vectors don't move; so
$\vec{V}$ stands for only $\vec{R}$ and $\vec{P}$ in this case.
This basically means that  $\vec{V}$ rotates around the fixed
vector $\vec{U}$ with an angular velocity whose magnitude is simply
$U$.

Constructing the solutions to the flows under $J$ and $L$
in terms of Cartesian 
components is cumbersome, so we will work with the magnitudes and
the directions of the vectors instead.
This paragraph assumes the reader is familiar with
the definitions of the frames $({i} {j} {k})$
and $({i'} {j'} {k'})$ which have been introduced with the help of
Fig.~\ref{fig:1} in Appendix~\ref{flow_under_SeffdL}.
Now in light of Eq.~\eqref{flow_rotation},
it is a simple matter to see that the 
equations for flow under $J$ and $L$ (or rather Eqs.~(21) and
(23) of Ref.~\cite{Tanay:2020gfb}) imply that 
\begin{itemize}
\item Under the flow of $J$ by an amount
 $\Delta \lambda$, the azimuthal angles of 
$\vec{R}, \vec{P}, \vec{S}_1$ and $\vec{S}_2$ 
in the inertial $(ijk)$ frame increase by $\Delta \lambda$.
The magnitudes of the vectors don't change.
\item  Under the flow of $L$ by an amount $\Delta \lambda$, 
the azimuthal angles of 
$\vec{R}$, and $\vec{P}$ in the non-inertial 
$(i'j'k')$ frame increase by $\Delta \lambda$, 
whereas the spin vectors don't move.
The magnitudes of the vectors don't change.
\end{itemize}
The flow under $J_z$ can be handled similarly.
With all the individual pieces now identified, it is now
straightforward, although lengthy to
find each standard phase space variable as an explicit function of the
angle variables $\theta^{i}$, on any invariant torus.

\subsection{Action-angle based solution at 1.5PN and higher PN orders}

Now there are two approaches to solving the real-time dynamics of the
system, i.e.\ a flow under $H$.  The approach by one us in
Ref.~\cite{cho2019analytic} was to directly 
integrate the differential equations,\footref{cho-footnote}
yielding a quasi-Keplerian parameterization.  Although this method is
direct, it seems quite difficult to extend this to
higher PN orders.  The second approach is the action-angle based one,
the subject of this paper.
All the angles have a
trivial real time evolution, each one increasing linearly with time
$\dot{\theta}^{i}=\omega^{i}(\vec{\mathcal{J}})$. After a certain time
$t$, $\theta^i$ has changed by $\omega^{i} t$, which we can compute.
So assuming that $\vec{\theta} (t= 0) = \vec{0}$, we can compute
the angles $\vec{\theta} (t)$ at any general time $t$, with the 
$\vec{\mathcal{J}}$ unchanged. Now the problem has become that of
computing  $(\vec{\mathcal{P}} , \vec{\mathcal{Q}})(t)$  given 
$( \vec{\mathcal{J}}, \vec{\theta}) (t)$, whose roadmap has been clearly 
laid out in  Sec.~\ref{angle-subsec}. This concludes our 
brief description of the action-angle based method of computing the
solution. This method has the 
advantage that evaluating the state of the system (or its derivatives,
as needed for computing gravitational waveforms) can be trivially
parallelized by evaluating each time independently.
Both the above solution methods have been implemented by us
in a public \textsc{Mathematica} package~\cite{Samanta:2022yfe}.

Moreover, our action-angle based solution allows for the possibility of using
non-degenerate perturbation theory~\cite{goldstein2013classical, jose} to
extend our solution to higher PN orders.
The procedure of Sec.~\ref{angle-subsec} will yield the standard phase-space variables
$(\vec{\mathcal{P}} , \vec{\mathcal{Q}})$ as explicit functions
of $(\vec{\mathcal{J}} , \vec{{\theta}})$.
This is exactly what is required for
computing perturbed action-angle variables at higher PN order with
canonical perturbation theory.
Higher-PN terms in the Hamiltonian are given in terms of
$(\vec{R},\vec{P},\vec{S}_1,\vec{S}_2)$, and one must transform
them to (unperturbed) action-angle variables to apply perturbation theory.
If successful, our method can be seen as the foundation
of closed-form solutions of BBHs with arbitrary masses, eccentricity, and spins to high PN orders
under the conservative Hamiltonian (excluding radiation-reaction for now).
This is in the same spirit as Damour and Deruelle's quasi-Keplerian solution method for non-spinning
BBHs given in Ref.~\cite{1985AIHS...43..107D}, 
which has been pushed to 4PN order recently~\cite{Cho:2021oai}.
We are also currently working to find the
 2PN action-angle based solution via canonical perturbation theory.

Note that we could not have applied non-degenerate perturbation theory
to a lower PN order (say 1PN) to arrive at 1.5PN or higher PN
action-angle variables, because the lower PN systems are degenerate in
the full phase space.  This is because the spin variables are not
dynamical until the 1.5PN order; so at lower orders, there are fewer
than four action variables and frequencies.\footnote{%
  There can be at most 4 different non-zero frequencies of this
  system irrespective of the PN order,
   since $H$ must be independent of $J_{z}$ to preserve SO(3)
  symmetry.}
At 1.5PN, the system becomes non-degenerate, and can be used as a
starting point for perturbing to higher order.
We therefore view our construction of the action-angle variables as
significant for finding closed-form solutions of the complicated
spin-precession dynamics of BBHs with arbitrary eccentricity, masses,
and spin.

\section{Summary and next steps}      \label{summary}

In this paper, we continue the integrability and action-angle variables study of the most general BBH system
(both components spinning in arbitrary directions,
 with arbitrary masses and eccentricity)
initiated in Ref.~\cite{Tanay:2020gfb}.
There, two of us presented four (out of five) actions at 1.5PN
and showed the integrable nature of the system at 2PN by constructing
two new 2PN perturbative constants of motion.
Here, we computed the remaining 
fifth action variable using a novel mathematical method of inventing
unmeasurable phase space variables.
We derived the leading order PN contribution to the fifth action,
which is a much shorter expression than the ``exact'' one.
We showed how to compute the fundamental frequencies of the system
without needing to write the Hamiltonian explicitly in terms of the actions.
Finally, we presented a recipe for computing the five angle variables implicitly,
by finding ($\vec{R},\vec{P},\vec{S}_1, \vec{S}_2$) as explicit functions of action-angle variables.
We leave deriving the full expressions to future work.
We also sketched how the 1.5PN action-angle variables can be used to construct solutions to the BBH system
at higher PN orders via canonical perturbation theory.

Typically, action-angle variables are found by separating the Hamilton-Jacobi (HJ)
equation~\cite{goldstein2013classical}, 
though we were able to work them out
without effecting such a separation.
Finally from this vantage point, we summarize the major
ingredients that went into our action-angle based solution for the PN BBHs:
(1) the classic Sommerfeld contour integration method for the Newtonian system,
which gave the Newtonian radial action long
ago~\cite{goldstein2013classical};
(2) its PN extension by Damour and Sch{\"a}fer~\cite{damour1988};
(3) the integration techniques worked out in the context of the 1.5PN Hamiltonian flow by one of us in Ref.~\cite{cho2019analytic};
and finally, (4) the method of extending the phase space by inventing
fictitious phase-space variables introduced in this paper.

A couple of extensions of the present work are possible in the near future.
Currently we are working on presenting our 1.5 PN action-angle-based solution
in a more concrete and consolidated form, 
as well as re-presenting the solution given in 
Ref.~\cite{Cho:2019brd} with 1PN terms included that were ignored in the
original work. We have developed a
public \textsc{Mathematica} package 
that implements these two solutions~\cite{Samanta:2022yfe}, as well as the one from
numerical integration.
This will prepare a solid base for pushing our action-angle-based
solution to 2PN.

Since the integrable nature (existence of action-angle variables)
 has already been shown in Ref.~\cite{Tanay:2020gfb},
constructing the 2PN action-angle variables (via canonical perturbation theory)
and an action-angle based solution
should be the next natural line of work.
Our group has already initiated the efforts in that direction.
With the motivation behind these action-angle variables study of the BBH systems being having closed-form solution
to the system, it would be an interesting challenge
 to incorporate the radiation-reaction effects at 2.5PN
into the to-be-constructed 2PN action-angle based solution. 
There is also hope that the
action-angle variables at 1.5PN can be used to re-present the 
effective one-body (EOB) approach to the spinning binary of Ref.~\cite{Damour2001}
(via a mapping of action variables between the one-body and the two-body pictures)
as was originally done for non-spinning binaries in Ref.~\cite{EOB}.
 Also, it would be interesting to try to
compare our action-angle and frequency results in the limit of extreme mass-ratios
with similar work on Kerr extreme mass-ratio inspirals (EMRIs)~\cite{Witzany2019} 
in some selected EMRI parameter 
space region where PN  approximation is also valid.
Comparison is also possible with the
 recently derived 
solution of EMRIs with spinning 
secondaries~\cite{Drummond:2022xej, Drummond:2022efc}.
Another line of effort could be
the task of building gravitational waveforms 
using the BBH solution presented in this paper;
Ref.~\cite{Konigsdorffer:2005sc}
may serve as one of the guides.

Lastly, there a possibility of a mathematically
oriented study of our novel method of introducing the unmeasurable,
fictitious variables to compute the fifth action. A few pertinent
questions along this line could be (1) Is there a way to compute the fifth action without introducing the 
fictitious variables? (2) Are there other situations (with other
topologically nontrivial symplectic manifolds)
where an otherwise intractable action computation 
can be made possible using this new method?
(3) What is the deeper geometrical reason that makes this method work?

\acknowledgements
We would like to thank Samuel Lisi for useful discussions.
We are also thankful to Rickmoy Samanta and
Jos\'e T. G\'alvez Ghersi 
 for a careful reading of this manuscript.
This work was partially supported by NSF CAREER Award
PHY--2047382.
G.C. is supported by the National Research Foundation (NRF) of Korea grant
 (No. 0409-20230090 and No. 0409-20230106), funded by the Ministry of Science and ICT (MSIT).

\appendix

\section{Evaluating the flow amounts $\Delta \lambda$'s}           \label{flow_under_SeffdL}

In this appendix, we will rely heavily on the
 methods of integration first presented in Ref.~\cite{cho2019analytic},
which integrated the evolution
 equations for flow under $H$, with the 1PN Hamiltonian terms omitted.
We first need to set up some vector 
bases before we can integrate the equations of motion. 
Fig.~\ref{fig:1} below displays
two sets of bases. The one in which the components of a vector will be assumed to be written in this paper
is the inertial triad $(ijk)$, unless stated otherwise.
Since derivatives of components of vectors depend on the basis, we mention here that
this $(ijk)$ triad is also the frame in which all the component derivatives of any general vector
will be assumed to be taken, unless stated otherwise.\footnote{%
  While the time derivative of a vector is a good geometric object,
  the time derivatives of basis components are not; see Ch.\ 4 of
  Ref.~\cite{goldstein2013classical}.}

\subsection{Evaluating $\Delta \lambda_{S_{\mathrm{eff}} \cdot L} $}       \label{mutual_angles_soln}

The evaluation of $\Delta \lambda_{S_{\mathrm{eff}} \cdot L} $ can happen only when
we can compute the mutual angles between $\vec{L}, \vec{S}_1$ and $\vec{S}_2$ as 
a function of the flow parameter under the flow of $\Eff$. Therefore, most of 
Appendix~\ref{mutual_angles_soln} deals with how 
to do this calculation and only towards the end 
we arrive at the expression of  $\Delta \lambda_{S_{\mathrm{eff}} \cdot L} $.

Under the flow of $\Eff$, a generic quantity $g$ evolves as $ dg/d\lambda = \pb{g, \Eff}  $
which implies the three evolution equations for the dot products between the three angular momenta under the flow of $\Eff$,
\begin{align}               \label{triple_cross_product}
 &   \frac{1}{\sigma_2} \frac{d ( \vec{L} \cdot \vec{S}_1)}{  d \lambda} =  - \frac{1}{\sigma_1} \frac{d ( \vec{L} \cdot \vec{S}_2)}{  d \lambda}       \nonumber \\  
   =  &     \frac{1}{(\sigma_1 - \sigma_2)} \frac{d ( \vec{S}_1 \cdot \vec{S}_2)}{  d \lambda}     = \vec{L}  \cdot (\vec{S}_1 \times \vec{S}_2) ,
\end{align}
which means that we can easily
construct three constants of motion (dependent on the five mutually commuting
constants as introduced before).  These are the differences between the three quantities
\begin{align}
  \left\{
  \frac{\vec{L}\cdot \vec{S}_{1}}{\sigma_{2}} \,,\
  -\frac{\vec{L}\cdot \vec{S}_{2}}{\sigma_{1}} \,,\
  \frac{\vec{S}_{1}\cdot\vec{S}_{2}}{\sigma_{1}-\sigma_{2}}
  \right\}\,,
\end{align}
whose $\lambda$ derivatives all agree, the triple product
$\tripleprod$.  Namely, these constants of motion are
\begin{align}
  \Delta_{1}   & = \frac{\vec{S}_{1}\cdot \vec{S}_{2}}{\sigma_{1}-\sigma_{2}} - \frac{\vec{L}\cdot \vec{S}_{1}}{\sigma_{2}}     \nonumber \\
  &=  \frac{1}{\sigma_{1}-\sigma_{2}}
  \left[
    \frac{1}{2}(J^{2}-L^{2}-S_{1}^{2}-S_{2}^{2}) - \frac{\LdotSeff}{\sigma_{2}}
  \right]                           \label{eq: Delta relation-1}
  \,,\\
  \Delta_{2}   &  = \frac{\vec{S}_{1}\cdot \vec{S}_{2}}{\sigma_{1}-\sigma_{2}} + \frac{\vec{L}\cdot \vec{S}_{2}}{\sigma_{1}}      \nonumber \\
   &= \frac{1}{\sigma_{1}-\sigma_{2}}
  \left[
    \frac{1}{2}(J^{2}-L^{2}-S_{1}^{2}-S_{2}^{2}) - \frac{\LdotSeff}{\sigma_{1}}
  \right]                          \label{eq: Delta relation-2}
  \,,\\
  \Delta_{21}   &  = \frac{\vec{L}\cdot\vec{S}_{1}}{\sigma_{2}} + \frac{\vec{L}\cdot\vec{S}_{2}}{\sigma_{1}}        \nonumber \\  
   &=   \frac{\LdotSeff}{\sigma_{1}\sigma_{2}}       \,.          \label{eq: Delta relation-3}
\end{align}
Stated differently, all this means that the three mutual angles
between $\vec{L}$, $\vec{S}_1$, and $\vec{S}_2$
satisfy linear relationships.  
With the understanding that hatted letters denote unit
vectors, if we define the mutual angles as
$\cos\kappa_{1}\equiv \hat{L}\cdot\hat{S}_{1}$,
$\cos\kappa_{2}\equiv \hat{L}\cdot\hat{S}_{2}$,
and
$\cos\gamma\equiv \hat{S}_{1}\cdot\hat{S}_{2}$,
their relations are
\begin{align}
\cos \gamma &=\Sigma_{1}+\frac{L}{S_{2}} \frac{\sigma_{1}-\sigma_{2}}{\sigma_{2}} \cos \kappa_{1}           \label{2nd_mutual_angle}    \\       
\cos \kappa_{2} &=\Sigma_{2}-\frac{\sigma_{1}S_{1}}{\sigma_{2}S_{2}}  \cos \kappa_{1}   ,                         \label{3rd_mutual_angle}
\end{align}
where
\begin{align} 
\Sigma_1 & =   \frac{(\sigma_1 - \sigma_2)\Delta_1}{S_1 S_2} =  \frac{\sigma_2(J^2 - L^2 -S_1^2 -S_2^2) - 2 \Eff}{2 \sigma_2 S_1 S_2 }  ,           \label{eq:Sigma_defined_1}    \\
\Sigma_2 & = \frac{ \Eff}{\sigma_2  L S_2}    =    \frac{\Delta_{21} \sigma_1 }{ L S_2}  .                                           \label{eq:Sigma_defined_2}
\end{align}

We will integrate the solution for
\begin{align}
  \label{eq:f_defined}
  f &\equiv \frac{\vec{S}_{1}\cdot\vec{S}_{2}}{\sigma_{1}-\sigma_{2}} =   \frac{ S_1 S_2 \cos \gamma  }{  \sigma_1 - \sigma_2} 
  \,, \\
  \label{eq:f_ODE}
  \frac{df}{d\lambda} &= \tripleprod \,,
\end{align}
which is the most symmetrical of the three dot products given above.
Thus if we have a solution for $f(\lambda)$, we automatically have
solutions for the three dot products,
\begin{align}
  \vec{S}_{1}\cdot \vec{S}_{2} &= (\sigma_{1}-\sigma_{2}) f    ,   \label{eq:threedots-1} \\
  \vec{L}\cdot \vec{S}_{1}        &= \sigma_{2} (f-\Delta_{1})    ,  \\
  \vec{L}\cdot \vec{S}_{2}        &= -\sigma_{1} (f-\Delta_{2})  .    \label{eq:threedots-3}
\end{align}

The triple product on the
RHS of Eq.~\eqref{eq:f_ODE} is the signed volume of the parallelepiped with ordered sides
$\vec{L}, \vec{S}_{1}, \vec{S}_{2}$.  In general, for a parallelepiped
with sides $\vec{  \mathcal{A}  }, \vec{  \mathcal{B}   }, \vec{   \mathcal{C}  }$, and dot products
\begin{align}
  \vec{   \mathcal{A}   }\cdot\vec{   \mathcal{B}  } &= \mathcal{A}  ~\mathcal{B} \cos\gamma'  ,  \\
  \vec{   \mathcal{A}  }\cdot\vec{   \mathcal{C}  } &= \mathcal{A}  ~ \mathcal{C} \cos\beta'  ,  \\
  \vec{   \mathcal{B} }\cdot\vec{ \mathcal{C} } &=  \mathcal{B}  ~ \mathcal{C}  \cos\alpha'  ,
\end{align}
a standard result from analytical geometry is that the signed volume of this parallelepiped can be written as
\begin{align}
  V                = \vec{\mathcal{A}} \cdot (\vec{\mathcal{B}}\times\vec{\mathcal{C}})      & =    \pm \mathcal{A}\mathcal{B}\mathcal{C}      \Big[  1  +2(\cos\alpha')(\cos\beta')(\cos\gamma')         \nonumber   \\
      &            - \cos^{2}\alpha'- \cos^{2}\beta'- \cos^{2}\gamma'    \Big]^{1/2}  ,
\end{align}
where the sign comes from the handedness of the $(\vec{\mathcal{A}}, \vec{\mathcal{B}},\vec{\mathcal{C}})$ triad.
The radicand is always non-negative.
As above in Eq.~\eqref{eq:threedots-1}-\eqref{eq:threedots-3}, we can rewrite all angles in terms of $f$.
We can then use this volume equation to express the evolution for $f$ as
\begin{align}
  \label{eq:df-by-dlambda}
  \frac{df}{d\lambda} = \pm \sqrt{P(f)}             
  \,,
\end{align}
where the cubic $P(f)  \geq   0$ and is given by
\begin{widetext}
\begin{align}
  P(f) &=
  L^{2}S_{1}^{2}S_{2}^{2}
  + 2 (\vec{L}\cdot\vec{S}_{1})(\vec{L}\cdot\vec{S}_{2})(\vec{S}_{1}\cdot\vec{S}_{2})
  - L^{2}(\vec{S}_{1}\cdot\vec{S}_{2})^{2}
  - S_{1}^{2}(\vec{L}\cdot\vec{S}_{2})^{2}
  - S_{2}^{2}(\vec{L}\cdot\vec{S}_{1})^{2}
  \\
    &=
  L^{2}S_{1}^{2}S_{2}^{2}
  - 2 \sigma_{1}\sigma_{2} (\sigma_{1}-\sigma_{2}) f (f-\Delta_{1}) (f-\Delta_{2})
  - L^{2}(\sigma_{1}-\sigma_{2})^{2} f^{2}
  - S_{2}^{2}\sigma_{2}^{2}(f-\Delta_{1})^{2}
  - S_{1}^{2}\sigma_{1}^{2} (f-\Delta_{2})^{2}                     \label{eq:cubic-1}
  \,.
\end{align}
\end{widetext}
This is a general cubic, which we will write as
\begin{align}
  P(f) = a_{3} f^{3} + a_{2} f^{2} + a_{1} f + a_{0} \,,        \label{P1}
\end{align}
with the coefficients
\begin{subequations}
  \label{eq:a-coeffs}
  \begin{align}
    \label{eq:a3}
    a_{3} &= 2 \sigma _1 \sigma _2 \left(\sigma _2-\sigma _1\right)
    \,,\\
    a_{2} &= 2 \left(\Delta _1+\Delta _2\right) \left(\sigma _1-\sigma _2\right) \sigma _1 \sigma _2
    -L^2 \left(\sigma _1-\sigma _2\right)^2     \nonumber \\  & -\sigma _1^2S_1^2 - \sigma _2^2 S_2^2
    \,,\\
    a_{1} &= 2 \left[ \sigma _1^2 S_1^2 \Delta _2+ \sigma _2^{2} S_2^2  \Delta _1 + \sigma _1\sigma _2 \Delta _1 \Delta _2 (\sigma _2-\sigma _1)  \right]
    \,,\\
    a_{0} &= L^2 S_{1}^{2}S_2^2 - \sigma _1^2 S_{1}^{2}\Delta _2^2 -  \sigma _2^2  S_2^2\Delta _1^2
    \,.
  \end{align}
\end{subequations}
It is important here to note the sign of $a_{3}$,
\begin{align}
  \text{sgn}(a_{3}) =
  \begin{cases}
    +1 \,, & m_{1} > m_{2} \,, \\
  \,\,\,\,   0 \,, & m_{1} = m_{2} \,, \\
    -1 \,, & m_{1} < m_{2} \,.
  \end{cases}
\end{align}
The fact that the cubic becomes undefined
when $m_{1}=m_{2}$
is the reason we treated the equal-mass case separately
towards the end of Sec.~\ref{sec:comp-fifth-action}.

Now we rewrite the cubic in terms of its roots,
\begin{align}
  P(f) = A (f-f_{1})(f-f_{2})(f-f_{3})      \label{P2}              
  \,,   
\end{align}
where $A=a_{3}$ is the leading term, and when all three roots are
real, we assume the ordering $f_{1} < f_{2}  < f_{3}$. In 
other words, we assume the roots to be real and simple.

For completeness, we state the roots in the trigonometric form.  The
cubic can be depressed by defining $g\equiv f + a_{2}/(3 a_{3})$ in terms of which
$P$ becomes $P = a_{3}(g^{3}+p g+q)$ with the coefficients
\begin{align}
  p &= \frac{3 a_{1}a_{3}-a_{2}^{2}}{3a_{3}^{2}} \,, &
  q &= \frac{2 a_{2}^{3} - 9 a_{1}a_{2}a_{3}+27a_{0}a_{3}^{2}}{27 a_{3}^{3}}
  \,.
\end{align}
When there are three real solutions, $p<0$, and the  argument to the
$\arccos$ below will be in $[-1,+1]$.  In terms of these depressed
coefficients, the trigonometric solutions for the $k=1,2,3$ roots are
\begin{align}
  \label{eq:f-trig-roots}
  f_{k} = - \frac{a_{2}}{3a_{3}} + 2\sqrt{\frac{-p}{3}} \cos
  \left[
    \frac{1}{3} \arccos
    \left(
      \frac{3q}{2p}\sqrt{\frac{-3}{p}}
    \right)+ \frac{2\pi k}{3}
  \right]
  \,.
\end{align}
This form yields the desired ordering $f_{1}  < f_{2}  < f_{3}$.

Whenever any two of the vectors $\{ \vec{L}, \vec{S}_{1}, \vec{S}_{2}
\}$ are collinear, the triple product on the RHS of
Eq.~\eqref{eq:f_ODE} vanishes.  A less drastic degeneracy is if two
roots coincide.  Here we will restrict ourselves to
the case of three simple roots. At the end of this subsection, we
will argue that the cubic has three real roots for the cases of
physical interest.  Since $P(f) > 0$, we have
\begin{align}               \label{mass_case}
  \begin{cases}
    f_{1} \le f \le f_{2} \,, & m_{1} > m_{2} \,, \\
    f_{2} \le f \le f_{3} \,, & m_{1} < m_{2} \,.
  \end{cases}
\end{align}
That is, $f$ will lie between the two roots where $P(f)>0$.
Without loss of generality we will take $m_{1}>m_{2}$ and handle only this case.

Since $P(f)$ is cubic, the ODE $df/d\lambda = \pm \sqrt{P(f)}$ can be
integrated analytically in terms of elliptic integrals (and their
inverses, elliptic functions). The behavior is typical: $f$ oscillates
between the two turning points $f_{1},f_{2}$ (when $m_{1}>m_{2}$).  We
can not integrate through the turning point using the first-order form
$df/d\lambda$ (it is not Lipschitz continuous there), but by taking a
derivative of Eq.~\eqref{eq:df-by-dlambda} to find
$d^{2}f/d\lambda^{2}$, we can see that the motion is regular at each
turning point.  At both turning points, the $\pm$ sign (the handedness
of the triad $(\vec{L}, \vec{S}_1, \vec{S}_2)$) must flip, so that $f$
oscillates between the two turning points.

Continuing further with Eq.~\eqref{eq:df-by-dlambda}, we write
\begin{align}
  \frac{df}{\sqrt{(f-f_{1})(f-f_{2})(f-f_{3})}} = \sqrt{A} d\lambda \,.
\end{align}
Reparameterize this integral via
\begin{align}
  \label{eq:f-param}
  f &= f_{1} + (f_{2}-f_{1})\sin^{2}\phi_{p}                   \\ 
  df &= 2 (f_{2}-f_{1}) \sin\phi_{p} \cos\phi_{p} \, d\phi_{p} \,.  \label{eq:f-param_2}
\end{align}
We define $\phi_p$ so it increases monotonically with $\lambda$ as
\begin{align}
  \label{eq:Legendreform}
  \frac{2 \, d\phi_{p}}{\sqrt{(f_{3}-f_{1})-(f_{2}-f_{1})\sin^{2}\phi_{p}}} = \sqrt{A} d\lambda \,.
\end{align}
Now factor out $(f_{3}-f_{1})$ from the radicand in the denominator to
give
\begin{align}
  \frac{d\phi_{p}}{\sqrt{1-  k^2 \sin^{2}\phi_{p}}} = \frac{1}{2}\sqrt{A(f_{3}-f_{1})} d\lambda \,,      \label{eq:incomp_ellip_int}
\end{align}
where we have defined the elliptic modulus
\begin{align}
  \label{eq:k}
  k  \equiv \sqrt{\frac{f_{2}-f_{1}}{f_{3}-f_{1}}}  .
\end{align}
Note that $0 < k < 1$, because of the ordering of the roots.
Equation \eqref{eq:incomp_ellip_int} can be integrated to give
\begin{align}
  \label{eq:F-phi-sol}
   u  - u_0   \equiv  F(\phi_{p} ,  k)  =  \frac{1}{2}\sqrt{A(f_{3}-f_{1})} ( \lambda-\lambda_{0}    )
  \,,
\end{align} 
where $F(\phi_{p}  , k)$ is the incomplete elliptic integral of the first kind 
defined as~\cite{NIST:DLMF, byrd2012handbook, whittaker1996course, lawden2013elliptic}
\begin{align}
F(\phi,k) \equiv \int_{0}^{\phi} \frac{\mathrm{d} \theta}{\sqrt{1-k^{2} \sin ^{2} \theta}}   .       \label{eq:ellip_int_defined}
\end{align}
In Eq.~\eqref{eq:F-phi-sol}, $\lambda_0$ is the initial value of the
flow parameter and  
\begin{align}
  u_0 \equiv  u(\lambda_0)   =    F\left(\arcsin \sqrt{\frac{f\left(\lambda_{0}\right)-f_{1}}{f_{2}-f_{1}}}   ,   k  \right)    .    \label{mutual-angle-sol-1}
\end{align} 
We can now rewrite the parameterization in terms of
$\sn$ and $\am$,
the Jacobi sine and amplitude functions~\cite{NIST:DLMF},
\begin{align}
  \sn(u,k)   \equiv \sin (\am (u,k)) \equiv   \sin \phi_p
  \,.
  \label{eq: ellip._sin_defined}
\end{align}
This turns our parameterization into
\begin{align}
f (\lambda)= f_1 +  (f_2 - f_1)    \sn^2 (u(\lambda) , k)   .          \label{mutual-angle-sol-2}
\end{align}
The solution for $f$ is thus given by Eq.~\eqref{mutual-angle-sol-2} accompanied by Eqs.~\eqref{eq:F-phi-sol} and \eqref{mutual-angle-sol-1}.

It now remains to generalize the 
solution for $f$ when $f$ at $\lambda = \lambda_0$
may be in any arbitrary initial state 
(such as $df/d \lambda < 0$ or $>0$)
and it can oscillate between $f_1$ and $f_2$
 any arbitrary number of times during the integration interval.
In this most general scenario, the solution is still given by 
Eq.~(\ref{mutual-angle-sol-2}), accompanied
by Eq.~(\ref{eq:F-phi-sol}) and a variant of (\ref{mutual-angle-sol-1}),
which reads
\begin{align}
  u_0 \equiv  u(\lambda_0)   =    F\left(\arcsin  \pm  \sqrt{\frac{f\left(\lambda_{0}\right)-f_{1}}{f_{2}-f_{1}}}   ,   k  \right)    ,    \label{mutual-angle-sol-111}
\end{align} 
where we use the $+$ sign if $(df/d \lambda) \vert_{\lambda_0} > 0$, and vice versa.

From this solution for $f(\lambda)$, we recover solutions for the
three dot products $\vec{S}_{1}\cdot\vec{S}_{2}$, $\vec{L} \cdot
\vec{S}_1$, and $\vec{L} \cdot \vec{S}_2$, by using Eqs.~\eqref{eq:threedots-1}-\eqref{eq:threedots-3}.
We also immediately get the
$\lambda$-period of the precession.  One precession cycle occurs when
$\phi_{p}$ goes from 0 to $\pi$, or when $f$ starts from $f_1$, goes to $f_2$
and then returns back to $f_1$ (see parameterization in
Eq.~\eqref{eq:f-param}). Integrating on this interval via Eq. \eqref{eq:F-phi-sol} gives the
equation for the $\lambda$-period of precession, which we call
$\Lambda$, in terms of the complete elliptic integral of the first kind $K(k) \equiv F(\pi/2 , k) = F(\pi , k)/2 $,
\begin{align}
    \Lambda      \sqrt{A(f_{3}-f_{1})}  =   2 F(\pi  , k ) = 4 K(k)                       \label{Lambda_sol}    .
\end{align}


Recall that our goal is to close a loop in the EPS by successively
flowing under $\Eff, J^2, L^2, S_1^2$, and $S_2^2$.
A necessary condition for the phase-space loop to close is that the
mutual angles between $\vec{L}, \vec{S}_1$, and $\vec{S}_2$ recur
at the end of the flow.
Since the flows under $ J^2, L^2, S_1^2$, and $S_2^2$ do not change
these mutual angles, we choose to flow under $\Eff$
by exactly the precession period,
\begin{align}
\Delta \lambda_{\Eff}    =  \Lambda    .     \label{lambda_1}
\end{align}
This flow under $\Eff$ is pictorially represented by the 
red $P Q$ curve in Fig.~\ref{fig:eps_sps_2}.

\figE

\figA

Now we try to address the issue of the nature
 of roots of the cubic $P(f)$ of Eq.~\eqref{P1}.
It is predicated on the nature of the cubic discriminant $D$,
with  $D> 0$ implying three real roots,  $D<0$ 
implying one real and two distinct complex roots,
and $D=0$ implying repeated roots.
The discriminant of the exact cubic $P(f)$ 
is too complicated for us
to investigate its sign analytically.
We rather choose to investigate the sign 
of its leading order PN contribution.
It is in the same spirit as the calculation 
of the leading PN order contribution
of $\mathcal{J}_5$ in Sec.~\ref{sec:lin_action}.
 We write $D$ 
in terms of $\vec{L}$, $\vec{S}_1$,
and $\vec{S}_2$ while attaching a formal power
 counting parameter $\epsilon$ to
both $\vec{S}_1$ and $\vec{S}_2$, for every factor
of $\epsilon$ signifies an extra 0.5PN order. 
Then series expand $D$ in $\epsilon$ and 
keep only the leading order term,
 which comes out to be
\begin{align}
D     & \sim  ~ 4 L^4 \left[ L^2 S_1^2 -   \left( \vec{L} \cdot \vec{S}_1  \right)^2   \right]  \left[ L^2 S_2^2 -   \left(\vec{L} \cdot \vec{S}_2  \right)^2   \right]  \nonumber \\
&   {}\times(\sigma_1 - \sigma_2)^6 \epsilon^{4} + \mathcal{O}(\epsilon^{5}) > 0
\,  ,
\end{align}
and this implies three real roots.
If both spins are aligned or anti-aligned with $\vec{L}$,
we will have repeated roots, and the spins will 
remain aligned or anti-aligned with $\vec{L}$
as the system evolves under the flows of $\Eff$ or $H$.
Aside from this special case, the above discussion suggests that the $D<0$
case of only one real root is disallowed.  This is also necessary on
physical grounds, as there must be two turning points for the mutual
angle variable $f$, otherwise $f$ would be unbounded.

\subsection{Evaluating $\Delta \lambda_{J^2} $}                         \label{restore-L}

After flowing under $\Eff$ by parameter $\Delta \lambda_{\Eff}$, 
the mutual angles between $\vec{L}, \vec{S}_1$, and $\vec{S}_2$
have recurred, but $\vec{L}$, $\vec{S}_1$, and $\vec{S}_2$ have not.
We now plan to flow under $J^2$ by  $\Delta \lambda_{J^2} $
so that $\vec{L}$ is restored; this restoration is a necessary
condition for closing the phase space loop.
To find the required amount of flow under $J^2$ so that $\vec{L}$ is restored,
we need to find the final state of $\vec{L}$ after flowing under $\Eff$ by $\Delta \lambda_{\Eff} $.
Instead of working with Cartesian components, we find it more
convenient to work with the polar and azimuthal angles of $\vec{L}$
in a new non-inertial frame that we now introduce.

At this point we introduce a non-inertial frame
 with $(i'j'k')$ axes
whose basis vectors are unit vectors along 
$  \vec{J} \times \vec{L} , \vec{L} \times (\vec{J} \times \vec{L} )$,
and $\vec{L}$ respectively, as depicted pictorially in Fig.~\ref{fig:1}.
Without loss of generality, we choose the 
$z$-axis of the $(ijk)$ frame to point along the $\vec{J}$ vector.
Now there are two angles to find: the polar $\theta_{JL}$,
where $\cos\theta_{JL} = \vec{J}\cdot\vec{L}/(J L)$, and an
azimuthal $\phi_{L}$.

Since we have already solved for the angles 
between $\vec{L}, \vec{S}_{1}$, and $\vec{S}_{2}$
in Appendix~\ref{mutual_angles_soln}, we have the angle $\theta_{JL}$ from
\begin{align}
  \vec{J}\cdot\vec{L} = J L \cos\theta_{JL} = L^{2} + \vec{S}_{1}\cdot\vec{L}+ \vec{S}_{2}\cdot\vec{L}
  \,.
\end{align}
This shows that $\theta_{JL}$ has recurred
after the $\Eff$ flow, because all the mutual angles between $\vec{L},
\vec{S}_1$, and $\vec{S}_2$ have. 
So, what remains to be tackled is the azimuthal angle $\phi_L$.
The inertial $(ijk)$ components of $\vec{L}$ are
\begin{align}
  \vec{L} = L ( \sin\theta_{JL} \cos\phi_{L}, \sin\theta_{JL} \sin\phi_{L}, \cos\theta_{JL} )
  \,,
\end{align}
and therefore it follows that
\begin{widetext}
\begin{align}
  \frac{d   \vec{L} }{d\lambda} = L
  \left(
    \cos\theta_{JL}\cos\phi_{L} \frac{d\theta_{JL}}{d\lambda} - \sin\theta_{JL} \sin\phi_{L} \frac{d\phi_{L}}{d\lambda},
    \cos\theta_{JL}\sin\phi_{L} \frac{d\theta_{JL}}{d\lambda} + \sin\theta_{JL} \cos\phi_{L} \frac{d\phi_{L}}{d\lambda},
    -\sin\theta_{JL} \frac{d\theta_{JL}}{d\lambda}
  \right)
  \,.
\end{align}
As mentioned in the beginning of Appendix~\ref{flow_under_SeffdL},
all vector derivatives are assumed to be taken in the inertial $(ijk)$
frame, unless stated otherwise.
With the aid of the instantaneous azimuthal direction vector given by
\begin{align}
\hat{\phi}=\frac{\vec{J} \times \vec{L}}{|\vec{J} \times \vec{L}|}=\frac{\vec{J} \times \vec{L}}{ J L \sin \theta_{J L}} =\frac{\hat{z} \times \vec{L}}{ L \sin \theta_{J L}}   \,    ,
\end{align}
we can extract $d\phi_{L}/d\lambda$ via an elementary result involving the dot product $\hat{\phi} \cdot  ({d \vec{L}}/{d\lambda} )  $
\begin{align}
  \hat{\phi}\cdot \frac{d \vec{L}}{d\lambda}  =L \sin\theta_{JL} \frac{d\phi_{L}}{d\lambda}
  \,.
\end{align}
This leads to
\begin{align}
  \frac{d \phi_{L}}{d\lambda}  = \frac{\hat{\phi}\cdot    (d\vec{L}/d\lambda )   }{ L \sin\theta_{JL}}
  = \frac{    \vec{J}\times\vec{L}     }{ J L^{2}\sin^{2}\theta_{JL}}     \cdot\frac{d\vec{L}}{d\lambda}
  \,.                          \label{phi_L_dot}
\end{align}
Now using $d\vec{L}/d\lambda = - d\vec{S}_{1}/d\lambda -
d\vec{S}_{2}/d\lambda $, and inserting the precession equations for
the two spins,
\begin{align}
  \frac{d  \phi_{L}  }{d\lambda}  &= \frac{1}{J L^{2}\sin^{2}\theta_{JL}} (\vec{J}\times\vec{L}) \cdot (\vec{S}_{\text{eff}}\times \vec{L})
  = \frac{  J  }{J^{2}L^{2}- (\vec{J}\cdot\vec{L})^{2}}
  \left[
    (\vec{J}\cdot\vec{S}_{\text{eff}})L^{2} - (\vec{J}\cdot\vec{L})(\LdotSeff)
  \right]
  \\
  &= \frac{  J  \left[
      (\sigma_{1} S_{1}^{2} + \sigma_{2} S_{2}^{2} + (\sigma_{1}+\sigma_{2})\vec{S}_{1}\cdot\vec{S}_{2})L^{2}
      - (\vec{S}_{1}\cdot\vec{L}+ \vec{S}_{2}\cdot\vec{L})(\LdotSeff)
    \right]}{J^{2}L^{2}- (L^{2} + \vec{S}_{1}\cdot\vec{L}+ \vec{S}_{2}\cdot\vec{L})^{2}}
  \,.
\end{align}

We see that everything on the RHS is given in terms of constants of
motion ($J, L, \LdotSeff$) and the inner products between the three angular 
momenta (which can be found from $f(\lambda)$ in the previous section).  Put everything
in terms of $f$ using
Eqs.~\eqref{eq:threedots-1}-\eqref{eq:threedots-3} and separate into
partial fractions,
\begin{align}
  \frac{d \phi_{L}}{d\lambda}  &= \frac{   J  \left[
      (\sigma_{1} S_{1}^{2} + \sigma_{2} S_{2}^{2} + (\sigma_{1}^{2}-\sigma_{2}^{2})f)L^{2}
      - (\sigma_{2}(f-\Delta_{1}) -\sigma_{1}(f-\Delta_{2}))(\LdotSeff)
    \right]}{J^{2}L^{2}- (L^{2} + \sigma_{2}(f-\Delta_{1}) -\sigma_{1}(f-\Delta_{2}))^{2}}
  \\
  &= \frac{B_{1}}{D_{1}-(\sigma_{1}-\sigma_{2})f} + \frac{B_{2}}{D_{2}-(\sigma_{1}-\sigma_{2})f}
  \,,                       \label{eq:phi_L_dot}
\end{align}
where we have defined
\begin{align}
  B_{1} &= \frac{1}{2}\left[(\LdotSeff+L^{2}(\sigma_{1}+\sigma_{2}))(J+L)
    +L\left(
      \sigma_{1}S_{1}^{2} + \sigma_{2}S_{2}^{2} + (\sigma_{1}+\sigma_{2})(\Delta_{2}\sigma_{1}-\Delta_{1}\sigma_{2})
    \right)\right] \,, \\
  B_{2} &= \frac{1}{2}\left[(\LdotSeff+L^{2}(\sigma_{1}+\sigma_{2}))(J-L)
    -L\left(
      \sigma_{1}S_{1}^{2} + \sigma_{2}S_{2}^{2} + (\sigma_{1}+\sigma_{2})(\Delta_{2}\sigma_{1}-\Delta_{1}\sigma_{2})
    \right)\right] \,, \\
  D_{1} &= L(L+J) + \Delta _2 \sigma _1 - \Delta _1 \sigma _2 \,, \\
  D_{2} &= L(L-J) + \Delta _2 \sigma _1 - \Delta _1 \sigma _2 \,.
\end{align}
So we need to be able to perform the two integrals (with $i=1,2$)
\begin{align}
  I_{i}  &  \equiv  \int \frac{B_{i}}{D_{i}-(\sigma_{1}-\sigma_{2})f} d\lambda = \int \frac{B_{i}}{D_{i}-(\sigma_{1}-\sigma_{2})f} \frac{d\lambda}{df} df       =  \int \frac{  \pm  B_{i}}{D_{i}-(\sigma_{1}-\sigma_{2})f} \frac{df}{\sqrt{A(f-f_{1})(f-f_{2})(f-f_{3})}} \,  ,    \label{eq:integral-I}
\end{align}
where the last equality is due to Eq.~(\ref{eq:df-by-dlambda}).
With these integrals, we will have
\begin{align}
  \int \frac{d\phi_{L}}{d\lambda} d\lambda = \phi_{L}(f) - \phi_{L,0} = I_{1} + I_{2} \,.
\end{align}
The integrals $I_{i}$ are another type of incomplete elliptic integral
(defined below).
Using the parameterization of Eqs.~\eqref{eq:f-param} and
\eqref{eq:f-param_2}, $I_i$ becomes
\begin{align}
  I_{i}(\lambda) &= \int^{\phi_{p}} \frac{B_{i}}{D_{i}-(\sigma_{1}-\sigma_{2})(f_{1}+(f_{2}-f_{1})\sin^{2}\phi_{p})} \frac{2d\phi_{p}}{\sqrt{A(f_{3}-f_{1})(1-k^{2}\sin^{2}\phi_{p})}}\\
  &=
  \frac{2 B_{i}}{\sqrt{A(f_{3}-f_{1})}} \frac{1}{D_{i}-f_{1}(\sigma_{1}-\sigma_{2})} \int^{\phi_{p}} \frac{1}{1-\alpha_{i}^{2}\sin^{2}\phi_{p}} \frac{d\phi_{p}}{\sqrt{1-k^{2}\sin^{2}\phi_{p}}}
  \,,
\end{align}
where we have defined
\begin{align}
  \alpha_{i}^{2} \equiv \frac{(\sigma_{1}-\sigma_{2})(f_{2}-f_{1})}{D_{i}-f_{1}(\sigma_{1}-\sigma_{2})}
  \,.
\end{align}
Thus we can identify the $I_{i}$'s in terms of the incomplete elliptic
integral of the third kind, which is defined as~\cite{NIST:DLMF}
\begin{align}
  \Pi(a, b, c) \equiv \int_{0}^{b} \frac{1}{\sqrt{1-c^{2} \sin ^{2} \theta}} \frac{d \theta}{1-a \sin ^{2} \theta}
  \,.
\end{align}
$I_i$ thus becomes
\begin{align}
  I_{i}( \lambda ) = \frac{2 B_{i}}{\sqrt{A(f_{3}-f_{1})}} \frac{ \Pi( \alpha_{i}^{2},\am (u(\lambda),k), k)}{D_{i}-f_{1}(\sigma_{1}-\sigma_{2})}
  \,,
\end{align}
and we get the solution for $\phi_{L}(\phi_{p})$
\begin{align}
  \phi_{L}(\lambda) - \phi_{L,0} = \frac{2}{\sqrt{A(f_{3}-f_{1})}}
  \left[
    \frac{B_{1} \Pi(\alpha_{1}^{2},   \phi_p  ,  k)}{D_{1}-f_{1}(\sigma_{1}-\sigma_{2})}
    +
    \frac{B_{2} \Pi( \alpha_{2}^{2},  \phi_p ,  k)}{D_{2}-f_{1}(\sigma_{1}-\sigma_{2})}
  \right]
  \,    .                                \label{eq:phi_L_soln}
\end{align}
Here $\phi_{L,0} $ is an integration constant to be determined by
inserting $\lambda=\lambda_{0}$ and $\phi_{L}=\phi_{L}(\lambda_{0})$
into the Eq.~\ref{eq:phi_L_soln}.

To close the loop, we need to know the angle $\Delta\phi_{L}$ that $\phi_{L}$ goes through under one
period of the precession cycle (when flowing under $\LdotSeff$), that is, when $\phi_{p}$ advances by
$\pi$.  This is given in terms of the \emph{complete} elliptic
integral of the third kind, $\Pi(\alpha^{2}, k)  \equiv  \Pi( \alpha^{2}, \pi/2, k)$ yielding 
\begin{align}
  \Delta\phi_{L} \equiv \phi_{L}(\lambda_0 + \Lambda) - \phi_{L}(\lambda_0) = \frac{4}{\sqrt{A(f_{3}-f_{1})}}
  \left[
    \frac{B_{1} \Pi(\alpha_{1}^{2}, k)}{D_{1}-f_{1}(\sigma_{1}-\sigma_{2})}
    +
    \frac{B_{2} \Pi(\alpha_{2}^{2}, k)}{D_{2}-f_{1}(\sigma_{1}-\sigma_{2})}
  \right]
  \,,            \label{eq:overflow_phi_L}
\end{align}
\end{widetext}
where we have used the fact that $\Pi( \alpha^{2}, \pi, k) = 2\Pi(\alpha^{2},k)$.

To negate this angular offset caused by flowing under $\Eff$ and
thereby closing the loop, we need to flow under $J^2$ by 
\begin{align}
\Delta \lambda_{J^{2}} =  -\frac{  \Delta \phi_L  }{2 J }      \label{lambda_2}   .
\end{align}
Note that this flow does not alter the mutual angles between
$\vec{L}, \vec{S}_1$, and $\vec{S}_2$, as is necessary to close the loop
in the phase space.
Now that the mutual angles within the triad
$(\vec{L},\vec{S}_{1},\vec{S}_{2})$ have recurred and the full
$\vec{L}$ vector has recurred,
the concern is if the spin vectors
have recurred or not.  The spin vectors are constrained not only by their mutual
angles with $\vec{L}$, but also $\vec{J}$.  
 Their angles with $\vec{J}$ are algebraically related
to the mutual angles that we have previously dealt with,
 e.g. $\vec{J}\cdot\vec{S}_{2} =
\vec{L}\cdot\vec{S}_{1} + \vec{S}_{1}\cdot\vec{S}_{2} + S_{2}^{2}$.

After the respective flows under $\Eff$ and $J^2$ by amounts indicated 
in Eqs.~\eqref{lambda_1} and \eqref{lambda_2},
all of these angle cosines between $\vec{L}, \vec{S}_1$, 
and $\vec{S}_2$ have recurred, which narrows things down to two
solutions: the original configuration for
$(\vec{L},\vec{S}_{1},\vec{S}_{2})$, and its reflection across the $J$-$L$
plane.  We can rule out the reflected solution with the following
observation.  The original configuration and its reflection have
opposite signs for the signed volume
$\vec{L}\cdot(\vec{S}_{1}\times\vec{S}_{2})$, and thus opposite signs
for the radical $\sqrt{P(f)}$ in
Eq.~\eqref{eq:df-by-dlambda}.
Now once we return back to the same point on the $f$ axis 
after flowing under $\Eff$, the handedness of the $(\vec{L}, 
\vec{S}_1, \vec{S}_2)$ triad is restored. This is because the handedness
must have flipped twice: first when $f$ touched $f_1$ and second when
it touched $f_2$.\footnote{%
There is also a complex-analytic interpretation. The function
$\sqrt{P(f)}$ is an analytic function on a Riemann surface of two
sheets.  The different signs of
$\vec{L}\cdot(\vec{S}_{1}\times\vec{S}_{2})$ correspond to being on
the two different sheets.  The solution is periodic after
completing a loop around both branch points, ending on the same
sheet where we started.
}
Therefore, after the flows by $\Eff$ and $J^{2}$
by the amounts specified in Eqs.~\eqref{lambda_1} and
\eqref{lambda_2}, each of the three
vectors $(\vec{L}, \vec{S}_{1},\vec{S}_{2})$ have recurred.
This second flow under $J^2$ 
is pictorially represented by the 
green $Q R$ curve in Fig.~\ref{fig:eps_sps_2}.

\subsection{Evaluating $\Delta \lambda_{L^2}$}                 \label{restore-RP}

After flowing under $\Eff$ and $J^2$,  all the three angular momenta
$\vec{L}, \vec{S}_1$, and $\vec{S}_2$ have recurred, but the orbital vectors ($\vec{R}, \vec{P}$)
  and fictitious vectors have not. We will now restore $\vec{R}$
and $\vec{P}$ by flowing under $L^2$ by  $\Delta \lambda_{L^2}$, to be
determined in this section.

Now, $\vec{R}$ has to be in the $i'j'$ plane because $\vec{R}  \perp \vec{L}$. 
Denote by $\phi$ the angle made by $\vec{R}$ with the $i'$ axis. The key point is that
after successively flowing under $\Eff$ by $\lambda_{\Eff}$, $ J^2$ by $\lambda_{J^2}$, and $L^2$
 by a certain amount $\lambda_{L^2}$ (to be calculated), if $\phi$ is restored, then
so are $\vec{R}$ and $\vec{P}$. This is because under these three flows,
$R, P$, and $\vec{R} \cdot \vec{P}$ do not change. Hence the restoration of $\phi$
after the above three flows by the stated amounts
restores both $\vec{R}$ and $\vec{P}$.

Our strategy is to compute $\phi$ under the flow of $\Eff$.
The flow under $J^2$ does not change the angle $\phi$, since
$J^{2}$ rigidly rotates all vectors together.
And in the end, we will undo the change to $\phi$ (caused by the $\Eff$ flow)
by flowing under $L^2$.

Under the flow of $\Eff$, we have
\begin{align}
\dot{ \vec{R}}  =  \pb{  \vec{R} , \Eff  }  =      \vec{S}_{\eff}  \times \vec{R}  .             \label{eq:R_dot_NIF}
\end{align}
To write the components of this equation in the $(i'j'k')$ frame,
 we need the components of
all the individual vectors involved in the same frame which are given by
\begin{align}
 \vec{R}      =
  \begin{bmatrix}
     R \cos \phi  \\
      R \sin \phi    \\
     0   
   \end{bmatrix}_{n}   ,    ~~~~
    \vec{L}
  & =~~~
       \begin{bmatrix}
      0     \\
        0     \\
    L 
   \end{bmatrix}_{n} 
          ,       \nonumber \\
    \vec{J}
         =
  \begin{bmatrix}
        0 \\
      J \sin \theta_{JL}  \\
     J \cos    \theta_{JL}  
   \end{bmatrix}_{n}   ,    ~~
      {\vec{S}}_1 
             &  =
   S_1  \begin{bmatrix}
     \sin \kappa_1  \cos \xi_1 \\
   \sin \kappa_1  \sin \xi_1   \\
    \cos \kappa_1   
   \end{bmatrix}_{n}
         ,       \nonumber \\
      \vec{S}_2
    &  =
     S_2  \begin{bmatrix}
        \sin \kappa_2  \cos \xi_2    \\
          \sin \kappa_2  \sin \xi_2        \\
       \cos \kappa_2   
   \end{bmatrix}_{n}             ,         \label{eq:vec_in_NIF}
\end{align}
where $\phi$ is the azimuthal angle of $\vec{R}$ in the $(i'j'k')$ frame. 
Here the letter `$n$' beside these columns indicates
that the components are in the $(i'j'k')$
frame, and $\xi_i$'s are the azimuthal angles
of $\vec{S}_i$ in this $(i'j'k')$ frame.

The Euler matrix $\tilde{\Lambda}$, which when multiplied with the column consisting of 
a vector's components in the inertial frame
 gives its components in the $(i'j'k')$ frame is
\begin{align}
\tilde{\Lambda}  =\left(\begin{array}{ccc}
- \sin \phi_{L} & \cos  \phi_{L} & 0 \\
-\cos \phi_{L} \cos \theta_{JL} & - \sin \phi_{L} \cos \theta_{JL} & \sin \theta_{JL} \\
\cos \phi_{L} \sin  \theta_{JL} &  \sin \phi_{L} \sin \theta_{JL} & \cos   \theta_{JL}
\end{array}\right)
\,.
\end{align}
Now we take the $\vec{R}$ in Eq.~(\ref{eq:vec_in_NIF}), evaluate its components in the
inertial frame using $\tilde{\Lambda}^{-1}$. We then differentiate
each of these components with respect to $\lambda$ (the flow parameter  under $\Eff$)
and transform these components back to the $(i'j'k')$ frame
 using $\tilde{\Lambda}$,
thus finally yielding the components (in the non-inertial frame) of
the derivative of $\vec{R}$.
The result comes out to be (keeping in mind that $dR/d\lambda = 0$)
\begin{gather}
  \dot{ \vec{R}}
      =
  \begin{bmatrix}
         - R \sin \phi   (    \dot{\phi_L}  \cos \theta_{JL}    + \dot{\phi} )     \\ 
             R \cos \phi   (    \dot{\phi_L}  \cos \theta_{JL}     + \dot{\phi} )        \\
      R  (   -  \dot{ \phi_L } \sin \theta_{JL}  \cos \phi     +    \dot{\theta}_{JL}  \sin \phi   ) 
   \end{bmatrix}_{n}   .                      \label{eq:R_dot_NIF_2}
\end{gather}
Plugging Eqs.~(\ref{eq:vec_in_NIF}) and (\ref{eq:R_dot_NIF_2}) in Eq.~(\ref{eq:R_dot_NIF}) and
using the first two components of the resulting matrix equation gives us 
\begin{align}
\frac{d \phi}{d \lambda}  =  \sigma_1 S_1  \cos \kappa_1  +  \sigma_2 S_2  \cos \kappa_2  - \cos \theta_{JL}  \frac{d \phi_L }{d \lambda}   \label{phi_dot_v2}
\end{align}
Note that what we need for Eq.~\eqref{eq:R_dot_NIF} is the
non-inertial-frame components of the frame-independent vector
$\dot{\vec{R}}$; not to be confused with the time derivatives of the
non-inertial-frame components of $\vec{R}$.

We digress a bit to write $\vec{J} = \vec{L} + \vec{S}_1 + \vec{S}_2$
in component form in the $(i'j'k')$ frame using
Eqs.~\eqref{eq:vec_in_NIF}. Only the third component is of interest to us, which reads
\begin{align}
  J \cos    \theta_{JL} =L+S_{1} \cos \kappa_{1}+S_{2} \cos \kappa_{2}
  \,.
  \label{eq:xi_2}
\end{align}
We use this equation for $\theta_{JL}$, and Eqs.~\eqref{eq:phi_L_dot}
for $d \phi_L/d \lambda$,
to write $d\phi/d\lambda$ in terms of $\kappa_1$, $\kappa_2$, and
$\gamma$.  Finally
using Eqs.~\eqref{eq:threedots-1}-\eqref{eq:threedots-3} to express
everything in terms of $f$, we get
\begin{widetext}
\begin{align}
  \frac{d \phi}{d \lambda}    =\frac{B_{1}}{D_{1}-\left(\sigma_{1}-\sigma_{2}\right) f}  - \frac{B_{2}}{D_{2}-\left(\sigma_{1}-\sigma_{2}\right) f}  - \frac{\Eff + (\Delta_1 - \Delta_2) \sigma_1 \sigma_2 + L^2 (\sigma_1 + \sigma_2)}{L} .                       \label{eq:phi_dot}
\end{align}
This is the equivalent of Eq.~\eqref{eq:phi_L_dot} for $d \phi_L/d \lambda$, and
therefore its solution can be found in a totally parallel way to
what led us to $\phi_{L}(\lambda)$ in Eq.~\eqref{eq:phi_L_soln}.
This gives us
\begin{align}
\phi(\lambda)-\phi_{0}=&\frac{2}{\sqrt{A\left(f_{3}-f_{1}\right)}}\left[\frac{B_{1} \Pi\left(\alpha_{1}^{2}, \phi_p , k\right)}{D_{1}-f_{1}\left(\sigma_{1}-\sigma_{2}\right)} - \frac{B_{2} \Pi\left(\alpha_{2}^{2},  \phi_p , k\right)}{D_{2}-f_{1}\left(\sigma_{1}-\sigma_{2}\right)}\right]
\nn\\
&{}- \left( S_{\mathrm{eff}} \cdot L+\left(\Delta_{1}-\Delta_{2}\right) \sigma_{1} \sigma_{2}+L^{2}\left(\sigma_{1}+\sigma_{2}\right) \right)  \frac{(\lambda - \lambda_0 ) }{L}
\,,
\label{eq:phi_soln}
\end{align}
where again the integration constant $\phi_{0}$ is determined by
inserting $\lambda=\lambda_{0}$ and $\phi=\phi(\lambda_{0})$ into this
equation.

The angle $\Delta\phi$ that $\phi$ goes through under one
period of the precession cycle when flowing under $\LdotSeff$,
is given in a similar manner as we arrived at Eq.~(\ref{eq:overflow_phi_L}). We get
\begin{align}
  \Delta\phi \equiv \phi(\lambda_0 + \Lambda) - \phi(\lambda_0)  & = \frac{4}{\sqrt{A(f_{3}-f_{1})}}
  \left[
    \frac{B_{1} \Pi(\alpha_{1}^{2}, k)}{D_{1}-f_{1}(\sigma_{1}-\sigma_{2})}
    -
    \frac{B_{2} \Pi(\alpha_{2}^{2}, k)}{D_{2}-f_{1}(\sigma_{1}-\sigma_{2})}
  \right]        \nonumber   \\
  &      -  \left( S_{\mathrm{eff}} \cdot L+\left(\Delta_{1}-\Delta_{2}\right) \sigma_{1} \sigma_{2}+L^{2}\left(\sigma_{1}+\sigma_{2}\right) \right) \frac{ \Lambda}{L}      \label{eq:Delta_phi_soln}
  \,.
\end{align}
\end{widetext}

To negate this angular offset caused by flowing under $\Eff$, we need to flow under $L^2$ by
\begin{align}
\Delta \lambda_{L^{2}} =  -  \frac{\Delta \phi}{2 L } .              \label{lambda_3}
\end{align}
Note that this flow does not change any of the three angular momenta
$\vec{L}, \vec{S}_1$, or $\vec{S}_2$, which is necessary
for closing the loop in the phase space.
This third flow under $L^2$ 
is pictorially represented by the 
blue $RS$ curve in Fig.~\ref{fig:eps_sps_2}.

\figB

\subsection{Evaluating $\Delta \lambda_{S_1^2} $ and $\Delta \lambda_{S_2^2} $}    \label{sec:subspin_ang._offset}

Once we have made sure that  $\vec{R}, \vec{P}, \vec{S}_1, \vec{S}_2$ (and hence also $\vec{L}$) have been
restored by successively flowing under $\Eff, J^2$, and $L^2$ by $\Delta \lambda_{\Eff} , 
\Delta \lambda_{J^2}$, and $\Delta \lambda_{L^2}$ respectively, now is the time to restore the fictitious vectors
$\vec{R}_{1/2}$ and $\vec{P}_{1/2}$. The strategy and calculations are analogous to the ones for $\vec{R}$ and $\vec{P}$,
so we won't explicate them in full detail. We will show the basic roadmap and the final results.

For the purposes of these calculations,
the relevant figure is Fig.~\ref{fig:2}, which shows a second non-inertial frame $(i''j''k'')$
adapted to $\vec{S}_1$. Its axes point along $\vec{J} \times \vec{S}_1 , \vec{S}_1 \times (\vec{J} \times \vec{S}_1 )$
and $\vec{S}_1$, respectively. We also use this figure to introduce the definitions
of the azimuthal angle $\phi_{S_1}$ and polar angle $\theta_{JS_1}$ pictorially.
Also, just like $\phi$ was the angle between $\vec{R}$ and the $i'$ axis in Appendix~\ref{restore-RP}, 
we define $\phi_1$ to be the angle between $\vec{R}_1$ and the $i''$ axis, with the understanding that 
$\vec{R}_1$  lies in the $i''j''$ plane.

 As far as the fictitious variables of the first black hole are
 concerned,
 just like in Appendices~\ref{restore-L} and \ref{restore-RP}, 
 all we have to worry about is to restore the change in
 $\phi_1$ which the $\Eff$ flow (by $\lambda_{\Eff}$) brings about, for doing so would imply that 
 both $\vec{R}_1$ and $\vec{P}_1$ have been restored. The justifications are analogous to those presented in
Appendices~\ref{restore-L} and \ref{restore-RP} while dealing with the orbital sector. 
Now we proceed to compute the change in $\phi_1$ brought about by the $\Eff$ flow.

We denote components in the ($i''j''k''$) frame by using the subscript
$n2$.   In this frame we have
\begin{gather}
 \vec{J}
 =
  \begin{bmatrix}
    0 \\
   J \sin \theta_{JS_1}  \\
   J \cos   \theta_{JS_1}
   \end{bmatrix}_{n2}   ,    
   ~~~~~~\vec{S}_1
 =
  \begin{bmatrix}
    0 \\
   0   \\
   S_1   
   \end{bmatrix}_{n2}   .
\end{gather}
We also have 
\begin{gather}
 \vec{L}
 =L
  \begin{bmatrix}
    \sin \kappa_1  \cos \xi_3 \\ 
   \sin \kappa_1  \sin \xi_3   \\ 
    \cos \kappa_1
   \end{bmatrix}_{n2}   ,    
  \vec{S}_2
 =  S_2
  \begin{bmatrix}
    \sin \gamma  \cos \xi_4  \\
     \sin \gamma  \sin \xi_4    \\
    \cos \gamma
   \end{bmatrix}_{n2}.         \label{eq:L_vector_matrix}      
\end{gather}
Here $\xi_3$ and $\xi_4$ are the azimuthal angles of $\vec{L}$ and $\vec{S}_2$, respectively, in the $(i''j''k'')$ frame.   
We now write the $k''$ component of $\vec{J} \equiv \vec{L} +
\vec{S}_1 + \vec{S}_2$ as
\begin{align}           
  J \cos \theta_{JS_1}   & = S_1  +   L \cos \kappa_1                 +  S_2  \cos \gamma
  \,.
  \label{suppl_1}
\end{align}
The derivative of $\vec{S}_1$ along the flow of $\Eff$ is
\begin{align}
\dot{\vec{S}}_1 \equiv  \frac{d \vec{S}_1}{d \lambda}  =  \pb{\vec{S}_1 , \vec{S}_{\text{eff}} \cdot \vec{L}} =    \sigma_1 \vec{L} \times \vec{S}_1.       \label{S1_dot_diff_eqn}
\end{align}

The analog of $d\phi/d\lambda$ given in Eq.~\eqref{phi_L_dot} becomes
\begin{align}
  \frac{d \phi_{S_1}}{d\lambda}  =    \frac{    \vec{J}\times\vec{S}_1     }{ J S_1^{2}\sin^{2}\theta_{J S_1}}     \cdot\frac{d\vec{S}_1}{d\lambda}
  \,.                          \label{phi_S1_dot}
\end{align}
Using Eq.~\eqref{S1_dot_diff_eqn}, we can arrive at the analog of
$d\phi/d\lambda$ as a function of $f$ [Eq.~\eqref{eq:phi_L_dot}],
\begin{align}
  \frac{d \phi_{S_1}}{d\lambda}  =  J \sigma_2 + \frac{B_{1S1}}{D_{1S1}+\sigma_{1} f} + \frac{B_{2S1}}{D_{2S1}+ \sigma_{1}f}
  \,,                       \label{eq:phi_S1_dot}
\end{align}
where we have defined
\begin{widetext}
\begin{align}
  B_{1S1} &=       \frac{1}{2}    \left[  -S_1 \sigma_1   (L^2 - J S_1 + S_1^2 +  \Delta_2  \sigma_1  )  + (J-S_1)^2 S_1  \sigma_2  - (J - 2 S_1) \Delta_1 \sigma_1 \sigma_2  + (J-S_1) \Delta_1 \sigma_2^2  \right]  , \\
 B_{2S1} &=       \frac{1}{2}    \left[  S_1 \sigma_1   (L^2 + J S_1 + S_1^2 +  \Delta_2  \sigma_1  )  - (J+S_1)^2 S_1  \sigma_2  - (J + 2 S_1) \Delta_1 \sigma_1 \sigma_2  + (J+S_1) \Delta_1 \sigma_2^2  \right] , \\
  D_{1S1} &=  (S_1 - J) S_1  - \Delta_1 \sigma_2                        \,, \\
  D_{2S1} &=   (S_1 + J) S_1   - \Delta_1 \sigma_2                \,.
\end{align}

Analogous to matrix equations for $\vec{R}$ and $\dot{\vec{R}}$ in Eqs.~(\ref{eq:vec_in_NIF}) and (\ref{eq:R_dot_NIF_2}),
we can write $\vec{R}_1$ in component form as
\begin{align}
\vec{R}_1=\left(\begin{array}{c}
R_1 \cos \phi_{1} \\
R_1 \sin  \phi_{1}  \\
0
\end{array}\right)_{n2}   ,                            \label{eq:R1_matrix}
\end{align}
and its derivative as (keeping in mind that $dR_1/d\lambda = 0$ along the flow under $\Eff$)
\begin{align}
\dot{\vec{R}}_1=\left(\begin{array}{c}
 - R_1 \sin \phi_{1} \left(   \dot{ \phi}_{S1} \cos \theta_{J S_1}   + \dot{\phi}_{1}\right) \\
  R_1 \cos \phi_{1} \left(          \dot{ \phi}_{S_1} \cos \theta_{J S_1}                  +\dot{\phi}_{1}\right) \\
R_1 \left(    -  \dot{  \phi}_{S_1} \sin \theta_{JS_1} \cos \phi_{1} + \dot{\theta}_{JS_1} \sin \phi_{1} \right)
\end{array}\right)_{n2}                                                         \label{eq:R1_dot_matrix_1}
\end{align}

Also, along the flow under $\Eff$, ${\vec{R}}_1 $ evolves as
\begin{align} 
\dot{\vec{R}}_1  =   \sigma_1    \vec{L}   \times    \vec{R}_1    .                        \label{eq:R1_dot_matrix_2}
\end{align}  
Using Eqs. \eqref{eq:L_vector_matrix}, \eqref{eq:R1_matrix}, and \eqref{eq:R1_dot_matrix_1} to express Eq. \eqref{eq:R1_dot_matrix_2}
in component form and either the first or the second component of the equation when supplemented with Eqs.~(\ref{suppl_1})
and (\ref{eq:phi_S1_dot}) to eliminate $ \cos \theta_{JS_1}$ and $d \phi_{S_1}/d \lambda$
gives us $\dot{\phi}_1$.
We again write the partial fraction form (analogous to Eq.~\eqref{eq:phi_dot})
\begin{align}
\dot{\phi}_{1}   =      S_1 (\sigma_2 - \sigma_1  )  -  \left(\frac{B_{1S1}}{ D_{1S1} + \sigma_1 f  } - \frac{B_{2S1}}{  D_{2S1}  + \sigma_1  f  }\right).
\end{align} 
We have also used Eqs.~\eqref{2nd_mutual_angle}, \eqref{3rd_mutual_angle}, and \eqref{eq:f_defined}
to write the cosines of $\kappa_1,\kappa_2$, and $\gamma$ in terms of $f$.

Finally, in a way very similar to how $\Delta\phi$ in Eq.~\eqref{eq:Delta_phi_soln} was found, we find
the angle $\Delta\phi_1$ that $\phi_1$ goes through under one
period of the precession cycle when flowing under $\Eff$. We get 
\begin{align}
\Delta \phi_{1}=\frac{- 4}{\sqrt{A\left(f_{3}-f_{1}\right)}}\left[\frac{B_{1S1} \Pi\left(\alpha_{1S1}^{2}, k\right)}{D_{1S1}+ f_{1} \sigma_{1}} - \frac{B_{2S1} \Pi\left(\alpha_{2S1}^{2}, k\right)}{D_{2S1}+f_{1}\sigma_{1}}\right] + S_1 (\sigma_2 - \sigma_1) \Lambda  ,          \label{eq:delta_phi_S_soln_1}
\end{align}
where we have defined
\begin{align}
  \alpha_{iS1}^{2} \equiv \frac{  - \sigma_{1}(f_{2}-f_{1})}{D_{iS1} + f_{1} \sigma_{1}}
  \,.
\end{align}
To negate this angular offset brought about flowing under $\Eff$, we need to flow under $S_1^2$ by
\begin{align}
\Delta \lambda_{S_{1}^{2}} =  -  \frac{\Delta \phi_{1}}{2 S_1 },    \label{lambda_4}
\end{align}
This fourth flow under $S_1^2$ 
is pictorially represented by the 
black $S T$ curve in Fig.~\ref{fig:eps_sps_2}.

And finally, by performing similar calculations as above, 
we can see that $\Delta \lambda_{S_{2}^2}$ (the amount we need to flow under $S_2^2$) is 
 given by the following set of equations
\begin{align}
\Delta \lambda_{S_{2}^2}   =   -  \frac{\Delta \phi_{2}}{2 S_2 },       \label{lambda_5}
\end{align}
\begin{align}
\Delta \phi_{2}=\frac{- 4}{\sqrt{A\left(f_{3}-f_{1}\right)}}\left[\frac{B_{1S2} \Pi\left(\alpha_{1S2}^{2}, k\right)}{D_{1S2}+ f_{1} \sigma_{2}} - \frac{B_{2S2} \Pi\left(\alpha_{2S2}^{2}, k\right)}{D_{2S2}+f_{1}\sigma_{2}}\right] + S_2 (\sigma_1 - \sigma_2) \Lambda  ,          \label{eq:delta_phi_S_soln_2}
\end{align}
\begin{align}
  B_{1S2} &=       \frac{1}{2}    \left[  S_2 \sigma_2   (L^2 - J S_2 + S_2^2  -  \Delta_1  \sigma_2  )  - (J-S_2)^2 S_2  \sigma_1     
  - (J - 2 S_2) \Delta_2 \sigma_1 \sigma_2  
  + (J - S_2) \Delta_2 \sigma_1^2  \right]  ,  \\
 B_{2S2} &=       \frac{1}{2}    \left[  S_2 \sigma_2   (-L^2 - J S_2 - S_2^2 +  \Delta_1  \sigma_2  )  + (J+S_2)^2 S_2  \sigma_1  
 - (J + 2 S_2) \Delta_2 \sigma_1 \sigma_2 
  + (J + S_2 ) \Delta_2 \sigma_1^2  \right] , \\
  D_{1S2} &=  \left(J-S_{2}\right) S_{2}-\Delta_{2} \sigma_{1}                       \,, \\
  D_{2S2} &=  -\left(J+S_{2}\right) S_{2} -\Delta_{2} \sigma_{1}               \,  ,
\end{align}
\begin{align}
\alpha_{i S 2}^{2} \equiv \frac{-\sigma_{2}\left(f_{2}-f_{1}\right)}{D_{i S 2}+f_{1} \sigma_{2}}.
\end{align}
This final fifth flow under $S_2^2$ 
is pictorially represented by the 
orange $T P$ curve in Fig.~\ref{fig:eps_sps_2}.
Of course, this final set of flows under $S_1^2$ and $S_2^2$ 
do not disturb the already restored configurations of the other
variables such as $\vec{R}, \vec{P}, \vec{S}_1$, and $\vec{S}_2$.
We mention that it is not recommended to try to
 arrive at  $\Delta \lambda_{S_{2}^2} $ from $\Delta \lambda_{S_{1}^2} $ by a mere 
 label exchange $ 1 \leftrightarrow 2$ (signifying the exchange of the two black holes)
 because we have already introduced asymmetry in these labels when we assumed
 $m_1 > m_2$ in Appendix~\ref{mutual_angles_soln}.

Finally, although not required for the fifth action computation, we mention
as an aside that the result of the integration of Eq.~\eqref{eq:phi_S1_dot} is
\begin{align}
\begin{aligned}
\phi_{S1}(\lambda)-\phi_{S10} =  \frac{2}{\sqrt{A\left(f_3-f_1\right)}}\left[\frac{B_{1S1} \Pi\left(\alpha_{1S1}^2, \phi_p, k\right)}{D_{1S1} + f_1  \sigma_1 } + \frac{B_{2S1} \Pi\left(\alpha_{2S1}^2, \phi_p, k\right)}{D_{2S1} + f_1  \sigma_1}\right]  +   J \sigma_2 {\left(\lambda-\lambda_0\right)}   ,
\end{aligned}
\end{align}
\end{widetext}
where again the integration constant $\phi_{S10}$ is determined by
inserting $\lambda=\lambda_{0}$ and $\phi_{S1}=\phi_{S1}(\lambda_{0})$ into this
equation.

\section{Proof that $\pi_{\star}(\mathcal{J}_5)$ is an action in the SPS}             \label{SPS-action}

By construction, $\J$ is an action variable in the EPS (as per the loop-integral
defintion), but we also
need to show that its pushforward $\pi_{\star}(\J)$ is an action (as
per the loop-flow definition) in the SPS; see Sec.~\ref{sec:EPS}
for these two definitions.
The pushforward can be constructed since $\J$ is fiberwise
constant.  To show that $\pi_{\star}(\J)$ is an action, we need to
show that (i) flowing under $\pi_{\star}(\J)$ forms a closed loop, and
(ii) this flow by parameter $2\pi$ takes us around the loop exactly
once.

Condition (i) can be shown to be satisfied automatically.  
Since the loop-integral
definition of action implies the loop-flow definition,
flowing under $\J$ in the EPS
forms a loop. Call this loop $\gamma$ (shown in solid cyan in
Fig.~\ref{fig:eps_sps_2}). The image of this
loop $\pi(\gamma)$ (shown in dashed cyan in Fig.~\ref{fig:eps_sps_2}) is a loop
in the SPS.  Meanwhile, because of the compatibility of the PBs
(see Sec.~\ref{sec:comparing-eps-sps}), the pushforward of the Hamiltonian
vector field $\pi_{\star}(\vec{X}_{\J})$ is the Hamiltonian vector
field of the pushforward,
$\vec{X}_{\pi_{\star}(\J)}=\pi_{\star}(\vec{X}_{\J})$.  Therefore
flowing under $\vec{X}_{\pi_{\star}(\J)}$ forms a loop, namely the
image $\pi(\gamma)$.

 The second part follows from homotopy equivalence. 
 In Fig.~\ref{fig:eps_sps_2}, let $\gamma_{1}$ be the path
$PQRS$ in the EPS, which is not a loop.  However, its image
$\pi(\gamma_{1})$ (in dashed red-green-blue)
is a loop in the SPS. Recall from Appendix~\ref{flow_under_SeffdL} that
we constructed $\gamma_{1}$ using three successive flows 
(under $\Eff, J^2$ and $L^2$)
to bring the SPS coordinates back to their starting values, thereby
making exactly one loop in the SPS.
 Let $\gamma_{2}$ be the segment $STP$, which
is vertical in the EPS (it is contained in a single fiber); its image
$\pi(STP)$ is a single point.  Their composition is
$\gamma_{3}=\gamma_{2}\cdot \gamma_{1}$, where $\cdot$ is composition
of paths.  Composing with the projection,
\begin{align}
  \pi(\gamma_{3}) = \pi(\gamma_{2}\cdot\gamma_{1}) = \pi(\gamma_{2})\cdot\pi(\gamma_{1}) = \pi(\gamma_{1})
  \,.
\end{align}

Now, the Liouville-Arnold theorem is constructive, meaning that when
we find the action
$\J = \oint_{\gamma_{3}}
\mathcal{P}_{i}d\mathcal{Q}^{i}/ (2 \pi) $, it generates a flow ($\gamma$) in
the same homotopy class as the path we integrated over
($\gamma_{3}$). The two loops are homotopic, i.e.,
 $[\gamma]=[\gamma_{3}]$, where the notation $[\gamma_{i}]$ denotes the
 homotopy class of a map $\gamma_{i}$.
Since $\pi$ is a continuous map, the two images are also homotopic.
Therefore we also have the homotopy
\begin{align}
  [\pi(\gamma)] &= [\pi(\gamma_{3})] = [\pi(\gamma_{1})]    \label{proof_eqn}
  \,.
\end{align}
Therefore we conclude that $\pi(\gamma)$ goes around 
exactly once, just like $\pi(\gamma_{1})$,
being in the same homotopy class.

\section{Frequently occurring derivatives in frequency calculations}        \label{derivative calculations}

Here we present some common derivatives that arise in the computation
of frequencies in Eqs.~\eqref{freq_eqn}.  The most important ones are
the derivatives of the roots $f_{i}$ of the cubic $P$.  These roots
are implicit functions of the constants of motion,
$f_{i}=f_{i}(\vec{C})$, and the coefficients of the cubic depend
explicitly on the constants, $P = P(f;\vec{C})$.  Since $f_{i}$ is a
root,
\begin{align}
  0 = P(f_{i}(\vec{C}); \vec{C})
  \,,
\end{align}
and this identity is satisfied smoothly in $\vec{C}$, therefore
\begin{align}
  0 &= \frac{\pd}{\pd C_{j}} 
  \left[
    P(f_{i}(\vec{C}); \vec{C})
  \right] \,,\\
  0 &=
  P'(f_{i}) \frac{\pd f_{i}}{\pd C_{j}} + \frac{\pd P}{\pd C_{j}}\bigg|_{f=f_{i}}
  \,,
\end{align}
where we have expanded with the chain rule.  We can now easily solve
for the derivative of a root with respect to a constant of motion,
\begin{align}
  \label{root_der}
  \frac{\pd f_{i}}{\pd C_{j}} = - \frac{1}{P'(f_{i})} \frac{\pd P}{\pd C_{j}}\bigg|_{f=f_{i}}
  \,.
\end{align}
Here $P'(f) = \pd P/\pd f$ is the quadratic
\begin{align}
  P'(f) = 3a_{3}f^{2} + 2a_{2}f+a_{1}
  \,,
\end{align}
where the coefficients are given in Eq.~\eqref{eq:a-coeffs}.
The denominator $P'(f_{i})$ only vanishes if $f_{i}$ is a multiple
root, which only happens if there is no precession.  Notice that
all the polynomials $\pd P/\pd C_{j}$ are also quadratics, since
the leading coefficient $a_{3}$ in Eq.~\eqref{eq:a3} does not depend
on any constants of motion.  We present these explicitly below.

Taking the derivative of $\mathcal{J}_5$ in Eq.~\eqref{eq:action_B}
requires applying the product rule and chain rule many times.  We need
the derivatives of the $\Delta \lambda$'s from Eqs.~\eqref{lambda_1},
\eqref{lambda_2}, \eqref{lambda_3}, \eqref{lambda_4}, and
\eqref{lambda_5}, which involve the quantities $f_i, B_i, D_i, B_{iS},
D_{iS}$ and various elliptic integrals.
Derivatives of $f_i$'s have already been discussed above and those of $B_i, D_i, B_{iS}, D_{iS}$
are not too hard to compute. Derivatives of the elliptic integrals via the application of the
chain rule can be written in terms of derivatives of their arguments:
$\alpha_i, \alpha_{iS}$ and $k$.
Derivatives of the first two can be written in terms of the
derivatives of $f_i, D_i$ and $D_{iS}$,
whereas the derivative of $k$ [Eq.~\eqref{eq:k}] simplifies to
\begin{align}    \label{k_der}
  \frac{d k}{ d C_i}  &=
  \frac{ {  - (f_2 -f_3)^2  \frac{\pd P}{\pd C_i}\bigg|_{f=f_1}   + (1 \rightarrow 2 \rightarrow 3) + (1 \rightarrow 3 \rightarrow 2)}}
  {2 k A (f_1-f_3)^2 (f_1 - f_2)(f_1 - f_3)(f_2 - f_3)}
  \,.
\end{align}

The $\partial P/\partial C_i$ polynomials occur in both Eqs.~\eqref{root_der} and \eqref{k_der}.
use of the expression of $P$ as given in Eq. \eqref{P2} is to be made to compute it. 
The non-zero $\partial P/\partial C_i$'s are the quadratic polynomials
\begin{widetext}
  \begin{align}
    \frac{\pd P}{\pd L} &= 2 L \Big[
    -(\sigma_{1}^{2}+\sigma_{2}^{2})f^{2}
    +
    \left(2~ \delta \sigma  \, \sigma_{1}\sigma_{2} \mathcal{G} - \delta\sigma^{-1}[ (\sigma_{1}+\sigma_{2}) \Eff + S_{1}^{2}\sigma_{1}^{2} +  S_{2}^{2}\sigma_{2}^{2}] \right) f \nn\\
    & \qquad+
    \mathcal{G}(S_{1}^{2}\sigma_{1}^{2}+S_{2}^{2}\sigma_{2}^{2})
    -
    \delta\sigma^{-2}  (S_{1}^{2}\sigma_{1}+S_{2}^{2}\sigma_{2})\Eff  + S_{1}^{2}S_{2}^{2} \Big] \,,\\
    \frac{\pd P}{\pd J}   & = 2J \Big[
    2\sigma_{1}\sigma_{2} f^{2}
    -
    \left(2 \delta \sigma  \, \sigma_{1}\sigma_{2} \mathcal{G} - \delta\sigma^{-1}[ (\sigma_{1}+\sigma_{2}) \Eff + S_{1}^{2}\sigma_{1}^{2} +  S_{2}^{2}\sigma_{2}^{2}] \right) f \nn\\
    & \qquad
    - \mathcal{G}(S_{1}^{2}\sigma_{1}^{2}+S_{2}^{2}\sigma_{2}^{2})
    +
    \delta\sigma^{-2}  (S_{1}^{2}\sigma_{1}+S_{2}^{2}\sigma_{2})\Eff \Big] \,,\\
    \frac{\pd P}{\pd \Eff}   & =
    2   \Big[  - (\sigma_{1}+\sigma_{2}) f^{2} + \Big[ \delta\sigma(\sigma_{1}+\sigma_{2})\mathcal{G}
    -   \delta\sigma^{-1}(2 \Eff + S_1^2\sigma_1 + S_2^2\sigma_2) \Big] f
     \nn\\
     & \qquad +
    \mathcal{G} (S_{1}^{2}\sigma_{1}+S_{2}^{2}\sigma_{2})
    - \delta\sigma^{-2}(S_{1}^{2}+S_{2}^{2})\Eff   \Big]
    \,,
\end{align}
where we have used the shorthands
\begin{align}
\delta \sigma  & = \sigma_1 - \sigma_2  ,  \\
\mathcal{G}  & =    \frac{ J^2 -L^2 -S_1^2 - S_2^2}{2 \, \delta \sigma^2}
\,.
\end{align}
The last piece are the derivatives of the complete elliptic integrals of the first and 
third kinds with respect to their arguments.  By differentiating their
integral definitions, the derivatives are expressible again as
elliptic integrals~\cite{NIST:DLMF},
\begin{align}
\frac{\mathrm{d}}{\mathrm{d} k} K(k)  &  =  \frac{E(k)}{k\left(1-k^{2}\right)}-\frac{K(k)}{k}    ,   \\
\frac{\partial \Pi(n, k)}{\partial n}   &  =   \frac{1}{2\left(k^{2}-n\right)(n-1)}\left(E(k)+\frac{1}{n}\left(k^{2}-n\right) K(k)+\frac{1}{n}\left(n^{2}-k^{2}\right) \Pi(n, k)\right)   ,    \\
\frac{\partial \Pi(n, k)}{\partial k}  &  =    \frac{k}{n-k^{2}}\left(\frac{E(k)}{k^{2}-1}+\Pi(n, k)\right).
\end{align}
\end{widetext}

\section{Refining the definition of PN integrability in Ref.~\cite{Tanay:2020gfb}}     \label{fix_definition}

The definition of PN integrability was first provided in Sec.~IV-A of Ref.~\cite{Tanay:2020gfb}
which was later refined in Sec.~IV-D, for it had some shortcomings. According to the refined definition,
we have $q$PN perturbative integrability in a $2n$-dimensional
phase space when we have $n$ independent phase-space
functions (including the $(q+1/2)$PN Hamiltonian) which are in
mutual involution up to at least $q$PN order.
One shortcoming in regard to even this refined
definition has come to our notice which we attempt to point out and fix in this appendix.

In Ref.~\cite{Tanay:2020gfb}, we noted that
$\widetilde{L^2}+c_{1}S_1^2 h/c^2 + c_{2}S_2^2 h/c^2$ was in mutual
perturbative involution with all the other phase space constants, for
arbitrary real coefficients $c_{1},c_{2}$, where $\widetilde{L^2}$ was
given by Eqs.~(50) and (53), and $h$ was defined in Eq.~(52) of
Ref.~\cite{Tanay:2020gfb}.  Similarly, for any real coefficients
$c_{3},c_{4},c_{5}$, the combination
$\widetilde{\Eff} + c_{3}S_1^2 h/c^2 + c_{4}S_1^2 h/c^2 +
c_{5}\vec{S}_1 \cdot \vec{S}_2/c^2$ was in mutual perturbative
involution with the other phase space constants.
It is important to note that the free terms with coefficients $c_{i}$ are
at the same PN order that we are keeping, and that they are not simply
built out of other constants of motion. With our previous definition
of PN integrability, this seems to suggest far more than $n$
independent functions in mutual perturbative involution.
This is in stark contrast with exact integrability scenario 
where one cannot have more than $n$ independent functions
in mutual involution on a $2n$ dimensional phase space. Clearly, something is wrong.

Another way to look at this problem is to realize that for 2PN integrability,
if we enumerate the required $n=5$ commuting constants by including
the 2.5PN Hamiltonian, $J^2$, $J_z$,
$\widetilde{L^2}$ and
$\widetilde{L^2}+c_{1}S_1^2 h/c^2 + c_{2}S_2^2 h/c^2$,
the latter two quantities will coincide
 in the PN limit $1/c \rightarrow 0$, thereby leaving
us with only four independent quantities in exact mutual involution,
 whereas the requisite number is 5 (both for PN
perturbative and exact integrability).
This means that the $1/c \rightarrow 0$ limit of the requisite number $n$ of
quantities in PN mutual involution (required for PN integrability) 
may not be enough for exact integrability (in the $1/c \rightarrow 0$ limit), which is bizarre.
The definition of PN integrability clearly needs a fix.

To fix the definition, we add one more demand:~the $n$ independent phase-space
functions (including the $(q+1/2)$PN Hamiltonian) must be such that in the limit $1/c \rightarrow 0$,
they reduce to $n$ independent phase-space
functions in exact mutual involution. As per this new definition,
we can't count $\widetilde{L^2}$ and
$\widetilde{L^2}+c_{1}S_1^2 h/c^2 + c_{2}S_2^2 h/c^2$
simultaneously in our list of independent functions in mutual
involution.  This remedies the aforementioned problems with the definition of PN integrability. It's easy to see that
the BBH system is still 2PN integrable per this revised definition of PN integrability since
$\widetilde{L^2}$ and $\widetilde{\Eff}$ reduce to $L^2$ and $\Eff$ in the $1/c \rightarrow 0$ limit,
which exactly mutually commute and are independent of each other.


\bibliography{pn_v2}

\end{document}


\title{Erratum: 
Action-angle variables of a binary black hole with arbitrary eccentricity, spins, and masses at 1.5 post-Newtonian order
 [Phys. Rev. D 107, 103040 (2023),
 arXiv: 2110.15351(v3)]}

\newcommand{\UMiss}{\affiliation{Department of Physics and Astronomy, The University of Mississippi, University, MS 38677, USA}}

\newcommand{\LUX}{\affiliation{LUX, Observatoire de Paris, Université PSL, Sorbonne Université, CNRS, 92190 Meudon, France}}

\newcommand{\utec}{\affiliation{
Universidad de Ingenieria y Tecnologia – UTEC, Jr. Medrano Silva 165 - Barranco 15063, Lima, Peru
}}

\author{Sashwat Tanay\,\orcidlink{0000-0002-2964-7102}}
\email{stanay@utec.edu.pe}
\LUX
\UMiss
\utec
\author{Leo C.~Stein\,\orcidlink{0000-0001-7559-9597}}
\email{lcstein@olemiss.edu}
\UMiss
\author{Gihyuk Cho\,\orcidlink{0000-0001-8813-270X}}
\affiliation{Department of Physics and Astronomy, Seoul National University, Seoul 151-742, Korea}

\hypersetup{pdfauthor={Tanay, Cho, and Stein}}

\maketitle

This erratum fixes errors in the previous arXiv preprint version (v3)
in regard to calculating the leading 
post-Newtonian (PN) order contribution of the fifth action integral. Henceforth, in this
erratum, this
version v3 of the preprint will be referred
to as the ``original article''. 
The correct standalone article, with the errors fixed follows this erratum.

The sketch of the calculation is laid out in Sec.~V of the 
original article in Steps 1-5. The first correction pertains to
Step 2. Replace  ``As a first step, series expand the cubic expression of Eq.~(A21), and its roots, keeping terms up to 
${\cal{O}}(\epsilon^2)$. Expansion of the roots up to 
${\cal{O}}(\epsilon^2)$ is necessary because the turning points $f_1$ and $f_2$ coincide at lower orders.'' with
``As a first step, series expand the roots of the cubic expression of Eq.~(A21), keeping terms up to 
${\cal{O}}(\epsilon^2)$.''
The above correction basically tells the reader 
not to series expand the cubic expression
but only its roots
up to ${\cal{O}}(\epsilon^2)$. This is the main error in the 
article, from which other errors originate.

We now proceed to correct the expression of the fifth action's
leading PN order contribution (given by Eqs.~(47)-(49) in
the original article).
With the following new definitions,
\begin{widetext}
\begin{align}
    \mathcal{A} &= -J^2 + L^2 + S_1^2 + S_2^2 , \\
    \mathcal{B}_1 &= 2 S_{\text{eff}} \cdot L + \mathcal{A} \sigma_1, \\
    \mathcal{B}_2 &= 2 S_{\text{eff}} \cdot L + \mathcal{A} \sigma_2, \\
    \mathcal{C}_1 &= \sqrt{L^2 S_2^2 - \frac{\mathcal{B}_1^2}{4 (\sigma_1 - \sigma_2)^2}} ,\\
    \mathcal{C}_2 &= \sqrt{L^2 S_1^2 - \frac{\mathcal{B}_2^2}{4 (\sigma_1 - \sigma_2)^2}}, \\
    \mathcal{D}  & = 
\sqrt{-\left( \frac{-\mathcal{B}_1^2 - \mathcal{B}_2^2 - 4 (2 \mathcal{C}_1 \mathcal{C}_2 - L^2 (S_1^2 + S_2^2)) (\sigma_1 - \sigma_2)^2}{\mathcal{B}_1^2 + \mathcal{B}_2^2 - 4 (2 \mathcal{C}_1 \mathcal{C}_2 + L^2 (S_1^2 + S_2^2)) (\sigma_1 - \sigma_2)^2} \right)},
\end{align}
the corrected expression of the fifth action 
variable at the leading PN order becomes 
\begin{equation}
\begin{split}
    \mathcal{J}_5  & \sim  \frac{1}{4 L (\sigma_1 - \sigma_2) \sqrt{ 1 + \frac{4 \mathcal{C}_1 \mathcal{C}_2}{-2 \mathcal{C}_1 \mathcal{C}_2 - L^2 (S_1^2 + S_2^2) + \frac{\mathcal{B}_1^2 + \mathcal{B}_2^2}{4 (\sigma_1 - \sigma_2)^2}} } \left(\mathcal{B}_1^2 + \mathcal{B}_2^2 - 4 (2 \mathcal{C}_1 \mathcal{C}_2 + L^2 (S_1^2 + S_2^2)) (\sigma_1 - \sigma_2)^2\right)} \\
    &\quad \times \Bigg[ -\mathcal{B}_1^3 (\mathcal{D} - 1) - \mathcal{B}_2^3 (\mathcal{D} - 1) - \mathcal{B}_1^2 \mathcal{B}_2 (\mathcal{D} + 1) - \mathcal{B}_1 \mathcal{B}_2^2 (\mathcal{D} + 1) \\
    &\quad + 4 (\sigma_1 - \sigma_2)^2 \Big( \mathcal{B}_1 (2 \mathcal{C}_1 \mathcal{C}_2 \mathcal{D} + L^2 ((\mathcal{D} + 1) S_1^2 + (\mathcal{D} - 1) S_2^2)) + \mathcal{B}_2 (2 \mathcal{C}_1 \mathcal{C}_2 \mathcal{D} + L^2 ((\mathcal{D} - 1) S_1^2 + (\mathcal{D} + 1) S_2^2)) \Big) \Bigg].
\end{split}
\end{equation}
\end{widetext}
Just like in the original article,
because of the assumption $m_1 > m_2$,
this expression is not
manifestly symmetric  with respect to
the label exchange $1  \leftrightarrow   2$, 
due to the $(\sigma_1 - \sigma_2)$
term in the denominator. However,
replacing $(\sigma_1 - \sigma_2)$ with $-|\sigma_1 - \sigma_2|$
restores the symmetry.

In the end, we would also like to withdraw the claim made
in the first and the last paragraphs of Sec.~V, that the
expression of the leading PN order of the fifth action can be used
to write the 1.5PN order Hamiltonian explicitly in terms of 
all the action integrals. This is so because the corrected expression of $\mathcal{J}_5$
presented above
is more complicated than the 
incorrect one
presented in Eq.~(47) of
the original article. Hence, it is no longer
apparent that a quartic equation in 
$S_{\text{eff}} \cdot L$ (with other action variables
as parameters) can indeed be written.
In the original article,
this quartic equation was 
the bridge between $\mathcal{J}_5$ and
the Hamiltonian as an 
explicit function of the action variables.
However, this realization poses no hurdle to computing the 
frequencies of the system, and can be done by inverting a 
certain Jacobian matrix (see Sec.~VI of the original article).

Other results and conclusions of the original article remain
unaffected.